\documentclass{aa}  

\usepackage{graphicx}
\usepackage{txfonts}
\usepackage{natbib}
\usepackage{color}

\begin{document}

   \title{Chemical composition of giant stars in the open cluster IC\,4756\thanks{Based on observations collected at the European Organization for Astronomical Research in the Southern Hemisphere under ESO programme 085.D-0093(A).}}


   \author{Vilius Bagdonas
          \inst{1}
          \and
          Arnas Drazdauskas
          \inst{1}
          \and
          Gra\v{z}ina Tautvai\v{s}ien\.{e}
          \inst{1}
          \and
          Rodolfo Smiljanic
          \inst{2}
          \and
          Yuriy Chorniy 
          \inst{1}
          }

   \institute{Institute of Theoretical Physics and Astronomy, Vilnius University, Saul\.{e}tekio al. 3, 10257 Vilnius, Lithuania\\
         \and
             Nicolaus Copernicus Astronomical Center, Polish Academy of Sciences, Bartycka 18, 00-716, Warsaw, Poland\\
             }

   \date{Received 24 January 2018 / Accepted 30 March 2018}

 
  \abstract
   {Homogeneous investigations of red giant stars in open clusters contribute to studies of internal evolutionary mixing processes inside stars, which are reflected in abundances of mixing-sensitive chemical elements like carbon, nitrogen, and sodium, while $\alpha$- and neutron-capture element abundances are useful in tracing the Galactic chemical evolution.}
   {The main aim of this study is a comprehensive chemical analysis of red giant stars in the open cluster IC\,4756, including determinations of $\rm ^{12}C/^{13}C$ and C/N abundance ratios, and comparisons of the results with theoretical models of stellar and Galactic chemical evolution.}
   {We used a classical differential model atmosphere method to analyse high-resolution spectra obtained with the FEROS spectrograph on the 2.2~m MPG/ESO Telescope. The carbon, nitrogen, and oxygen abundances, $\rm ^{12}C/^{13}C$ ratios, and neutron-capture element abundances were determined using synthetic spectra, and the main atmospheric parameters and abundances of other chemical elements were determined from equivalent widths of spectral lines. 
   }
   {We have determined abundances of 23 chemical elements for 13 evolved stars and $\rm ^{12}C/^{13}C$ ratios for six stars of IC\,4756. The mean metallicity of this cluster, as determined from nine definite member stars, is very close to solar – [Fe/H] = $-0.02\pm 0.01$. Abundances of carbon, nitrogen, and sodium exhibit alterations caused by extra-mixing: the mean $\rm ^{12}C/^{13}C$ ratio is lowered to $19\pm1.4$, the C/N ratio is lowered to $0.79\pm0.05$, and the mean [Na/Fe] value, corrected for deviations from the local thermodynamical equilibrium encountered, is enhanced by $0.14\pm0.05$~dex. We compared our results to those by other authors with theoretical models.}
   {Comparison of the $\alpha$-element results with the theoretical models shows that they follow the thin disc $\alpha$-element trends. 
   Being relatively young ($\sim 800$~Myr), the open cluster IC\,4756 displays a moderate enrichment of $s$-process-dominated chemical elements compared to 
    the Galactic thin disc model and confirms the enrichment of $s$-process-dominated elements in young open clusters compared 
    to the older ones. The $r$-process-dominated element europium abundance agrees with the thin disc abundance. 
    From the comparison of our results for mixing-sensitive chemical elements and the theoretical models, we can see that the mean values of 
    $^{12}{\rm C}/^{13}{\rm C}$, C/N, and [Na/Fe] ratios lie between the model with only the thermohaline extra-mixing included and the model which also includes the rotation-induced mixing. The rotation was most probably smaller in the investigated IC\,4756 stars than 30\% of the critical rotation velocity when they were on the main sequence.
    }


   \maketitle


\section{Introduction}

Our Galaxy has always occupied an important place in the field of astrophysics. In today's era of space- and ground-based surveys like $Gaia$ \citep{Prusti16}, $Gaia$-ESO \citep{Gilmore12}, APOGEE \citep{Nidever12}, GALAH \citep{DeSilva15}, and RAVE \citep{Kunder17}, there is an abundance of data for studying 
the formation and evolution of the Milky Way. However, even massive surveys do not include every object, or every aspect of selected objects. That is why employing smaller telescopes and individual observations of selected targets is still important, providing valuable information that is otherwise lost due to various constraints that are often present in massive surveys. In this paper, we continue our investigation of red giant branch stars in open clusters from high-resolution spectra by analysing the open cluster IC\,4756. 

Stars in open clusters are known to be born from the same molecular cloud at roughly the same time and distance. Sharing the same primordial material, the open cluster stars have similar metallicity and chemical composition. New open clusters are constantly identified and the available sample of these stellar associations expands in variety of ages, heliocentric and galactocentric distances, and metallicities. 
Stars in open clusters have a significant advantage over the field stars. Precise ages that can be determined for open clusters allow studies of how the properties of the Galactic disc, such as chemical patterns for example, change with time, providing important constraints for Galactic modelling.
Furthermore, while members of open clusters have a common origin, they still have a different initial mass, which predetermines the lifetime of a star. For this reason, stars in open clusters are crucial tools in understanding how photospheric chemical composition varies due to stellar evolutionary effects.  

Every chemical element in stellar atmospheres provides valuable information. 
Some of them, such as carbon and nitrogen as investigated in this work, being susceptible to stellar evolutionary effects, show alterations of their observable abundances in photospheres of evolved stars. The standard stellar evolution model only predicts one mixing event on the red giant branch (RGB) - called the first dredge-up \citep{Iben65}, when the convective envelope expands and connects the inner layers containing the nuclear processed material and allows it to rise to the surface. However, as observational evidence suggests (\citealt{Smiljanic09, Mikolaitis10, Mikolaitis11a, Mikolaitis11b, Drazdauskas16, Drazdauskas16b, Tautvaisiene16} and references therein), after the luminosity bump on the RGB, there is a further decrease of $^{12}$C and increase of $^{14}$N in low-mass stars, which cannot be explained by standard stellar evolution models. For example, in stars of solar metallicity and about 2.5 solar mass, the carbon abundance may decrease by 0.3~dex, and the nitrogen abundance may increase by 0.3~dex (\citealt{Charbonnel10}). The exact causes of this effect are not yet fully understood and there are models which predict different extra-mixing effects depending on stellar turn-off mass and metallicity \citep{Chaname05, Charbonnel06, Cantiello10, Charbonnel10, Denissenkov10, Lagarde11, Lagarde12, Wachlin11, Angelou12, Lattanzio15}. 

In recent years there has been an increase in studies of carbon and nitrogen in field and open cluster stars. Together with theoretical models (\citealt{Charbonnel10, Lagarde12}) we try to constrain possible mechanisms governing the extra-mixing in evolved stars. In our work, in addition to carbon and nitrogen ratios, we look at carbon isotopic ratios ($\rm ^{12}C/^{13}C$), which are less susceptible to systematic errors in stellar atmospheric parameters and provide an even better insight into the extra-mixing phenomenon. 

 Abundances of other chemical elements like oxygen and $\alpha$- and iron-peak elements play an important role when trying to constrain theoretical models of Galactic chemical evolution. Oxygen and $\alpha$-elements are mainly produced via Type~II supernovae explosions and the timescales of elemental injection into the interstellar medium are short. Fe-peak elements, produced by Type Ia supernovae, on the other hand, have longer interstellar medium enrichment timescales \citep{Tinsley79, Matteucci86, Wyse88, Yong05}. As [Fe/H] and [$ \alpha $/Fe] steadily increase and decrease over time, respectively, the ratios of these elements are good age or place-of-birth indicators of specific objects of the Galaxy \citep{Matteucci92}. It is also established that the scatter of abundances of [$ \alpha $/Fe] at the same [Fe/H] value can help us distinguish between thin and thick discs \citep{Mikolaitis14, Masseron15}. 

Abundances of neutron-capture (n-capture) chemical elements also deserve further investigations. A recent study of low-mass stars by \cite{D'Orazi09} showcased an interesting phenomenon: the abundance of the slow neutron capture process ($s$-process) element barium seems to increase with decreasing cluster age. The barium abundances for young open clusters exhibited the enhancements up to $\sim$0.6 dex. While later studies confirmed this finding for barium, to the same or a lesser extent, the abundance enrichment patterns for other s-process elements in young open clusters are still debatable (e.g. \citealt{ Maiorca11, Yong12, D'Orazi12, Jacobson13, Mishenina13, Mishenina15, ReddyLambert15, D'Orazi17}). The suggested explanations for higher $s$-process element yields include higher production rates of these elements than previously thought by very old, low-mass AGB stars \citep[e.g.][]{Maiorca11}, additional contribution by the $i$-process \citep{Mishenina15}, or overestimation of barium abundances by a standard LTE abundance analysis \citep{Reddy17}. Overall, a comparison of n-capture chemical element abundances not only with age, but with other important parameters such as metallicity or galactocentric distance (e.g. \citealt{Jacobson13, Overbeek16}) provides  important information about stellar evolution and about the Galaxy as a whole. 

We compare our results with theoretical models, which depict aspects of both Galactic and stellar chemical evolution. Models by \cite{Magrini09}, \cite{Pagel97}, and \citet{Maiorca12} trace the Galactic evolution as a function of the galactocentric distance, metallicity, and age, respectively, whereas models by \cite{Lagarde12} provide insight into the stellar evolution and alterations of light elements in atmospheres of evolved stars.

\section{Cluster IC\,4756}

IC\,4756 is a relatively young open cluster in our Galaxy. Like most young open clusters, IC\,4756 is located in the inner part of the Galactic disc (the galactic coordinates \textit{l} = $36^\circ.381$, \textit{b} = $5^\circ.242$), at a galactocentric distance $R_{\rm gc}$ of around 8.1~kpc \citep{Gilroy89}. In order to separate cluster members from field stars, the first photometric study for this cluster was made by \citet{Kopff43}, and later by \citet{Alcaino65}, \citet{Seggewiss68}, and \citet{Herzog75}. There are several cluster age determinations: an age of 0.82~Gyr was derived by \citet{Alcaino65}, 0.8~Gyr by \citet{Gilroy89}, 0.79~Gyr by \citet{Salaris04}, $0.8\pm0.2$~Gyr by \citet{Pace10}, and most recently 0.89$\pm$0.07~Gyr was determined by \citet{Strassmeier15}.

\begin{figure}
   \centering
        \includegraphics[width=\columnwidth]{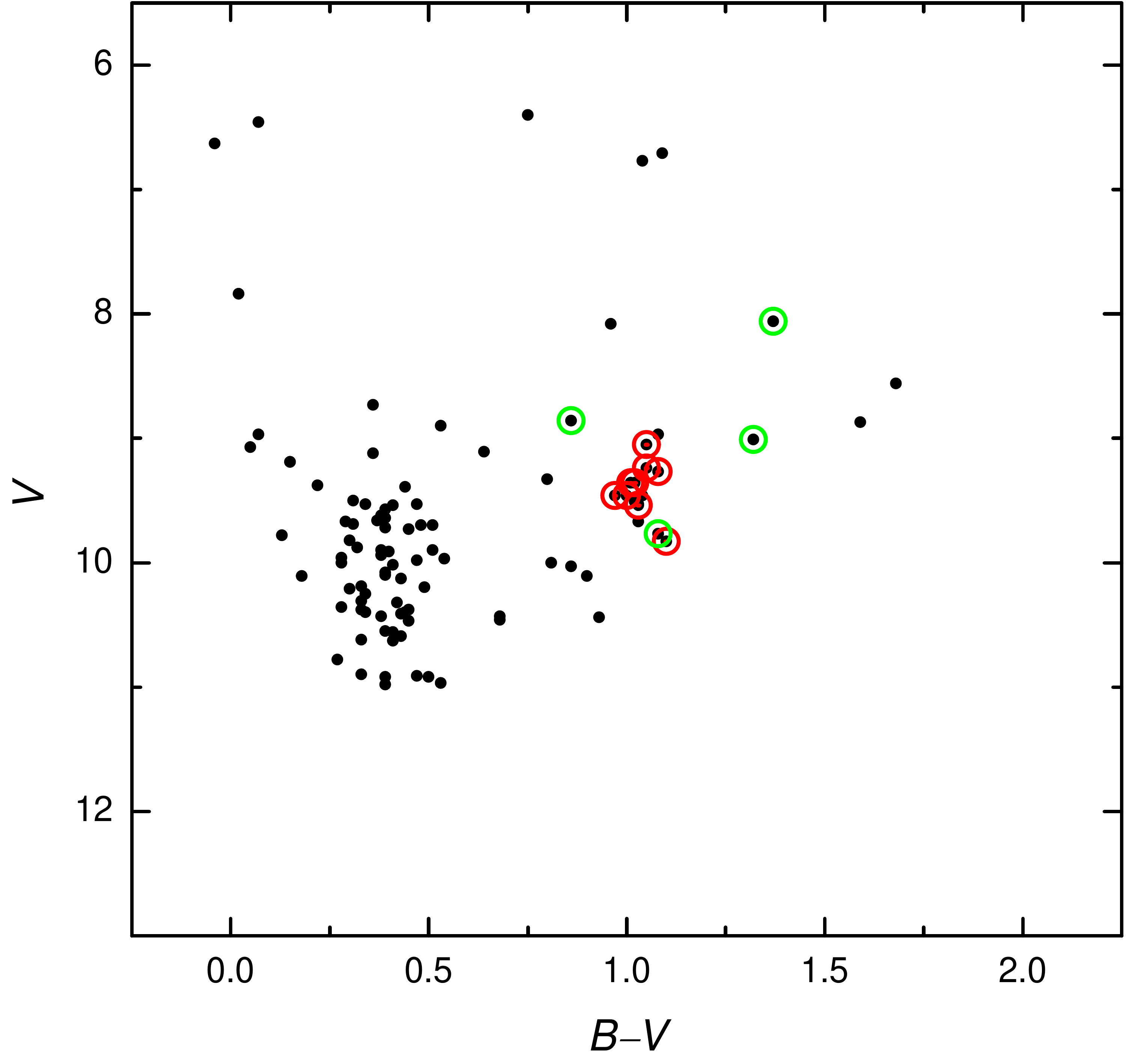}
   \caption{A colour-magnitude diagram of the open cluster IC\,4756. The $UBV$ photometry of the open cluster is from \cite{Alcaino65}. The stars investigated in this work are marked with circles. The green circles represent the stars with uncertain membership status.}
              \label{CMD}%
    \end{figure}

There are a few studies of this cluster in which metallicity was determined from photometry. \citet{Smith83}, using a $DDO$ cyanogen colour excess parameter, calculated the cluster metallicity [Fe/H]$ = 0.04\pm0.05$. The most recent comprehensive study of this cluster was made by \citet{Strassmeier15}. In their work, an averaged metallicity from several CMD diagrams was $-0.03\pm0.02$~dex and a combined metallicity from spectroscopy and photometry was $-0.08\pm0.06$~dex.

\citet{Gilroy89}, using high-resolution spectroscopy from seven giant stars, determined a cluster metallicity [Fe/H]$ = 0.04\pm0.07$. In a later study of this cluster performed by \citep{Luck94}, four giant stars gave a metallicity [Fe/H]$ = -0.03\pm0.05$. Slightly sub-solar metallicities of [Fe/H]$ = -0.22\pm0.12$ (8 stars) and [Fe/H]$ = -0.15\pm0.04$ (6 stars) were derived using moderate-resolution CCD spectra by \citet{Thogersen93} and \citet{Jacobson07}, respectively. A later study by \citet{Twarog97}, using  the $DDO$ photometry from \citet{Smith83} and data from \citet{Thogersen93}, provided the weighted average metallicity of $-0.06\pm0.10$~dex. The recent study by \citet{Smiljanic09} determined metallicities of $0.04\pm0.03$~dex. It is worth mentioning that \citet{Santos09}, using a  line-list by \citet{Sousa08}, determined metallicity of $0.02\pm0.02$~dex, while for the same stars a value of $\rm{[Fe/H]}=0.08\pm0.01$ was obtained using a line-list by \citet{Hekker07}.  The most recent entirely spectroscopic study of 12 giant stars in this cluster was published by \citet{Ting12}, in which a metallicity of $-0.01\pm0.1$~dex was determined. 

However, these works investigated relatively small samples of stars and were mostly dedicated to $\alpha$- and iron-peak chemical elements. A more comprehensive study of C, N, and O abundances is required, as these abundances were previously determined only by \citet{Smiljanic09} for three stars, and an upper value of oxygen abundance was evaluated also from three stars by \citet{Pace10}. The $\rm ^{12}C/^{13}C$ ratios for this cluster were also derived only by \citet{Gilroy89} and \citet{Smiljanic09}. A deeper analysis for the n-capture elements is also needed. The only determinations of abundances for these elements were performed by \citet{Luck94} (Ba, Y, Zr, Nd, and Eu abundances presented), by \citet{Maiorca11} (provided abundances of Y, Zr, La, Ce) and by \citet{Ting12} (abundances of only Ba determined). Thus, the deficiency of studies of neutron capture elements for this cluster and also its young age, make IC\,4756 the important object to analyse when trying to constrain $s$- and $r$-process element enrichment patterns in young ages. 

\section{Observations and method of analysis}

\subsection{Observations}
The spectra of our programme stars were observed using the bench-mounted, high-resolution astronomical \/{e}chelle spectrograph FEROS (Fiber-Fed Extended Range Optical Spectrograph) \citep{Kaufer99} at the 2.2~m MPG/ESO Telescope in La Silla, between  June 27 and 30, 2010. The FEROS covers a  whole visible range of $3500-9200$~\AA \,over 39 orders with the resolving power of about 48\,000. The FEROS Data Reduction System pipeline within MIDAS was used for spectral reductions. Depending on magnitudes of the observed objects, exposure times where chosen to achieve signal to noise ratios (S/N) higher than 120 at about 6400~\AA. The most luminous object was observed for 120 seconds, while others have observation times ranging from 300 to 600 seconds with a maximum of 900 s. A total of 13 stars in the cluster IC\,4756 were observed. We list them in Table~\ref{list_of_stars}. The identifications of stars were adopted from \citet{Kopff43}. 
    
\begin{table*}
        \centering
    \caption{A list of analysed stars and their parameters.}
    \label{list_of_stars}
    \begin{tabular}{lcccccccc}
    \hline
    \hline
    \noalign{\smallskip}
ID      &       R.A.(2000)&     DEC(2000)&      $V$     &       $B-V$   &       Date Obs.    &       Exp. time        &       Rad. vel.       &       S/N     \\
        &       deg &   deg  &  mag     &       mag     &               &       seconds &       $\rm km\,s^{-1}$     &   (at $\sim6400$~\AA)         \\
        \hline
        \hline
        \noalign{\smallskip}
12      & 278.9476      & +5.3381       &       9.54    &       1.03    &       2010-06-28      &       600     &       $-25.25$        &       $\sim$140       \\
14      & 278.9936      & +5.4167       &       8.86    &       0.86    &       2010-06-30      &       600     &       $-24.82$        &       $\sim$180    \\
28      & 279.1384      & +5.2119       &       9.01    &       1.32    &       2010-06-30      &       600     &       $-25.26$        &       $\sim$160       \\
38      & 279.2717      & +5.2921       &       9.83    &       1.10    &       2010-06-27      &       900     &       $-25.78$        &       $\sim$140       \\
42      & 279.3365      &  +5.8953      &       9.46    &       0.97    &       2010-06-30      &       600     &       $-24.92$        &       $\sim$140       \\
44      & 279.3762      & +5.2044       &       9.77    &       1.08    &       2010-06-30      &       900     &       $-25.00$        &       $\sim$140       \\
52      & 279.3992      &  +5.2605      &       8.06    &       1.37    &       2010-06-30      &       120     &       $-25.21$        &       $\sim$120       \\
69      & 279.5215      & +5.4094       &       9.24    &       1.05    &       2010-06-27      &       600     &       $-27.70$        &       $\sim$150       \\
81      & 279.5865      & +5.4340       &       9.46    &       1.00    &       2010-06-30      &       600     &       $-23.25$        &       $\sim$140       \\
101     & 279.6824      & +5.2389       &       9.36    &       1.02    &       2010-06-30      &       600     &       $-24.70$        &       $\sim$140       \\
109     & 279.7205      & +5.3379       &       9.05    &       1.05    &       2010-06-30      &       300     &       $-25.25$        &       $\sim$120\\
125     & 279.8245      & +5.2302       &       9.36    &       1.01    &       2010-06-30      &       600     &       $-24.00$        &       $\sim$140       \\
164     & 280.0771      & +5.3144       &       9.27    &       1.08    &       2010-06-30      &       600     &       $-25.51$        &       $\sim$150       \\
    \noalign{\smallskip}
    \hline
    \end{tabular}
    \tablefoot{The IDs are adopted from \citet{Kopff43}, the $UBV$ photometry are taken from \cite{Alcaino65}, the radial velocities as determined in this work. }
\end{table*}
    
 \begin{figure*}
\resizebox{\hsize}{!}{
\includegraphics[scale=0.3]{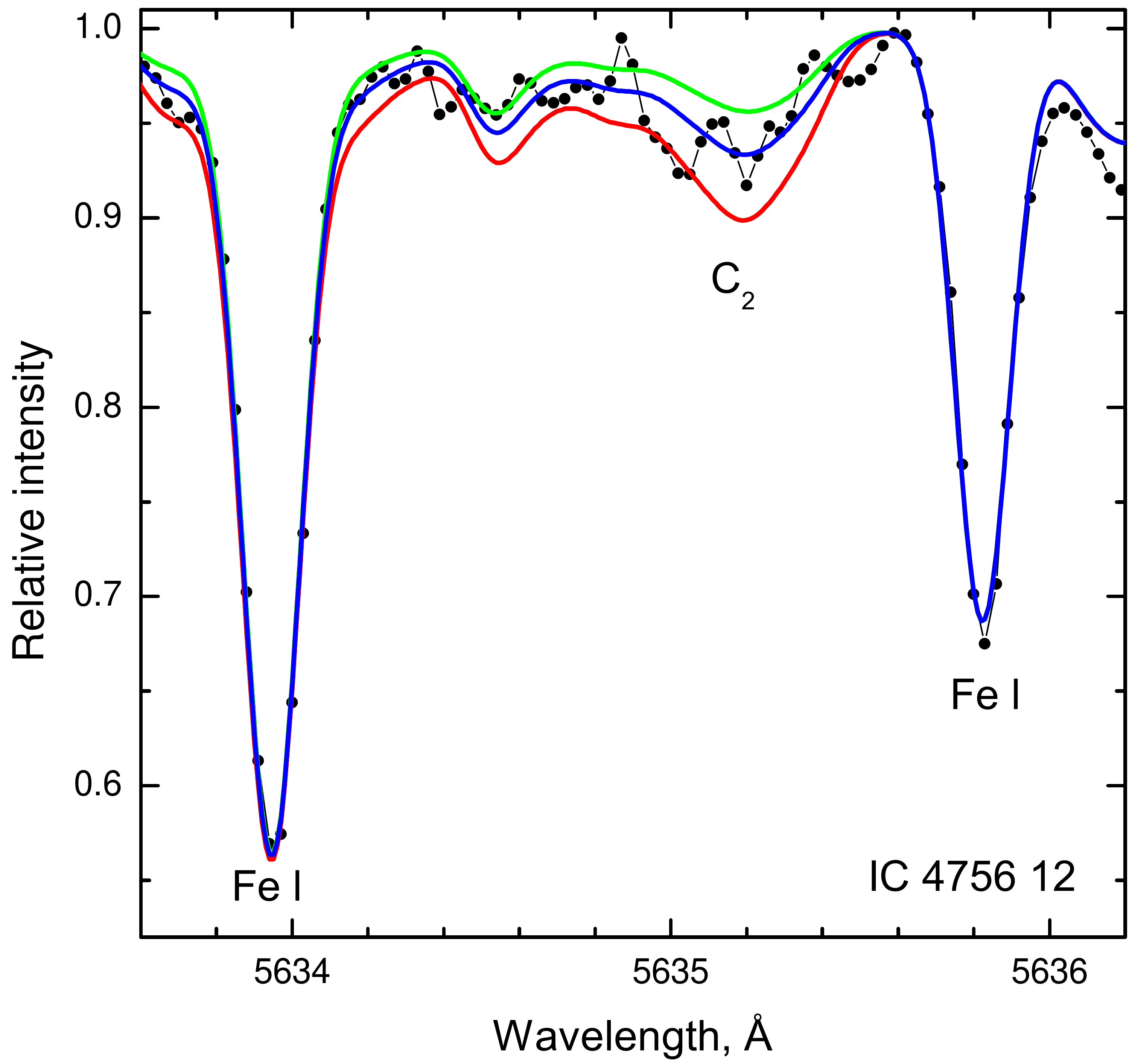}
\includegraphics[scale=0.3]{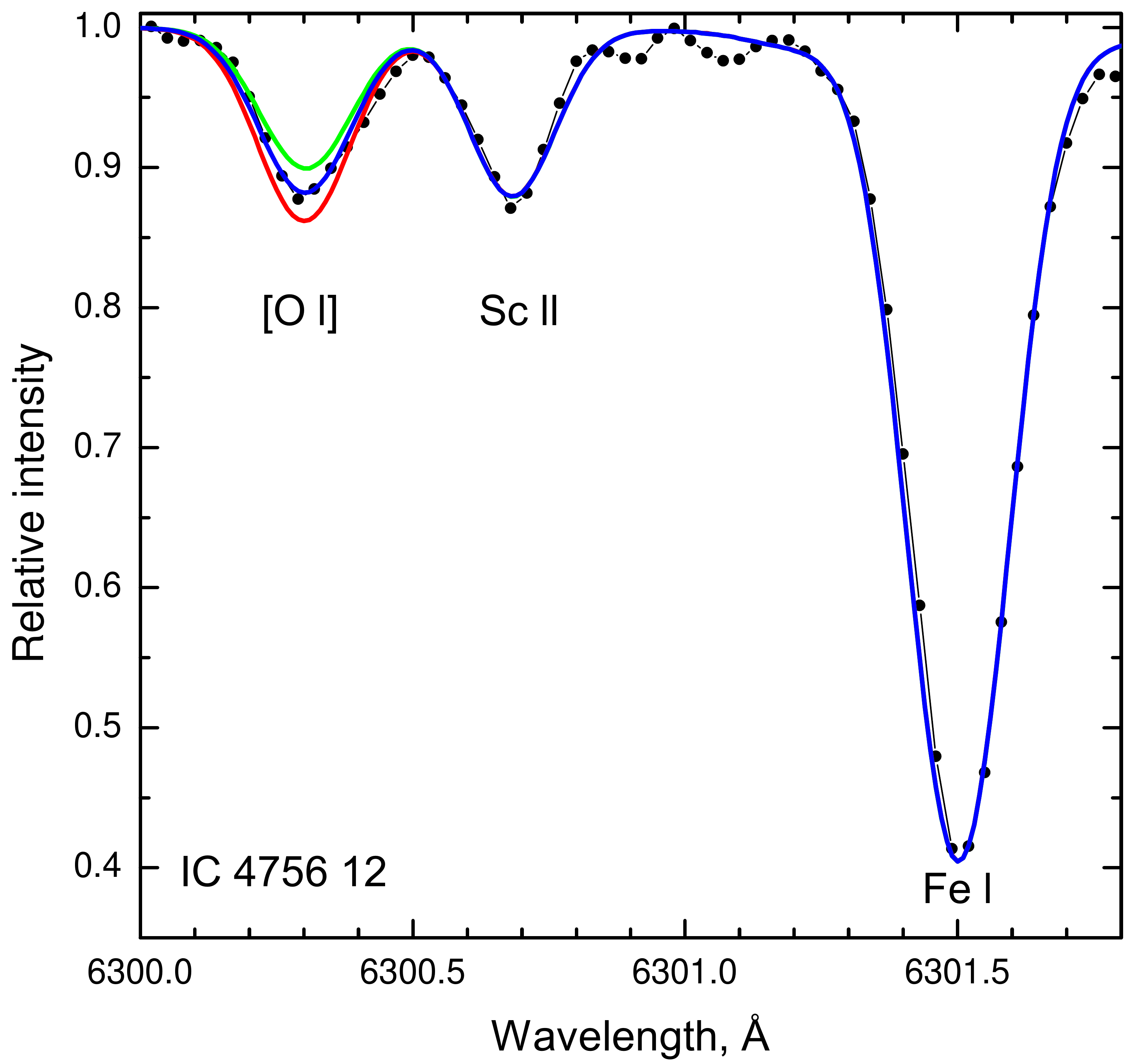}
\includegraphics[scale=0.3]{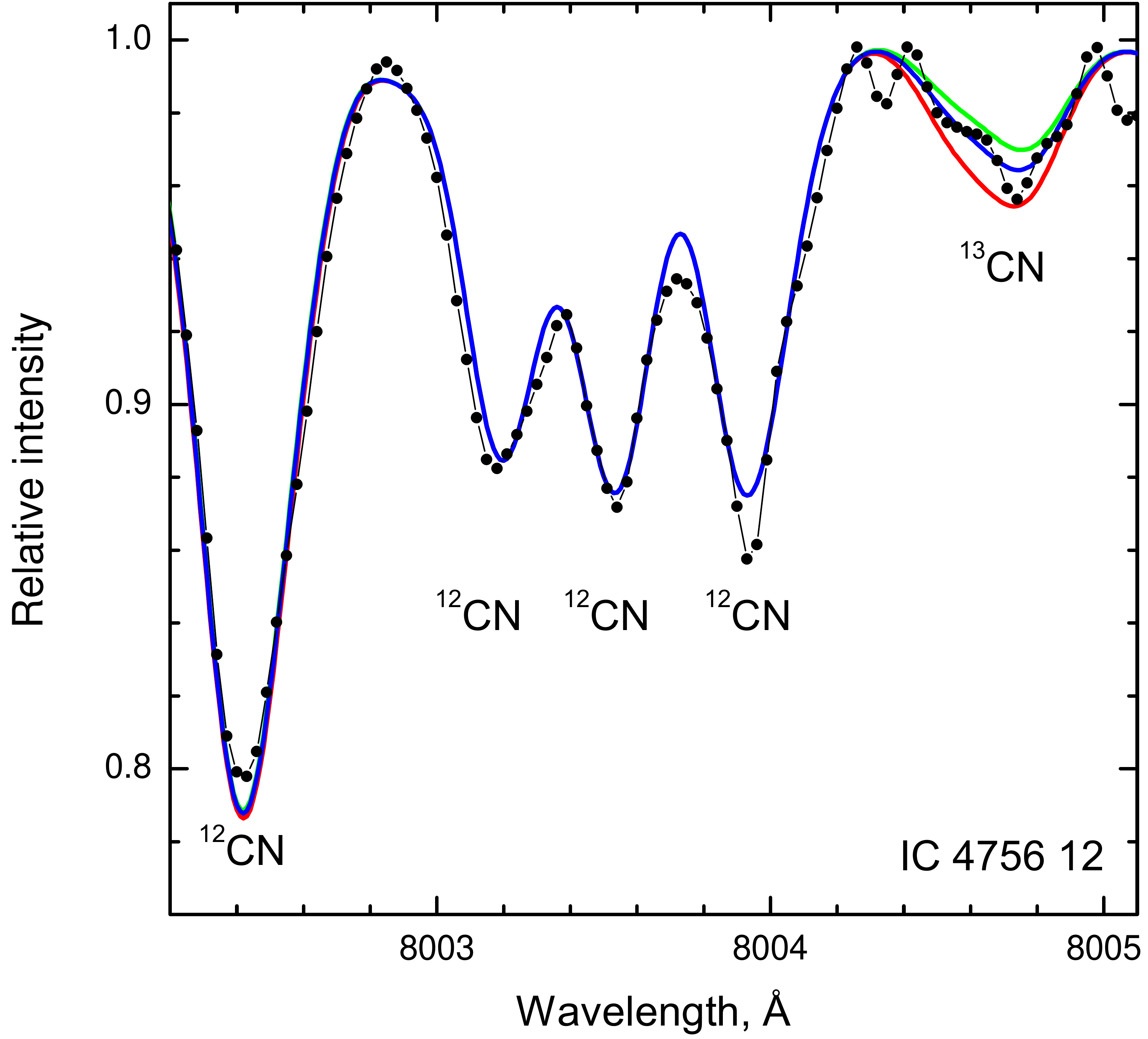}
}
\resizebox{\hsize}{!}{
\includegraphics[scale=0.3]{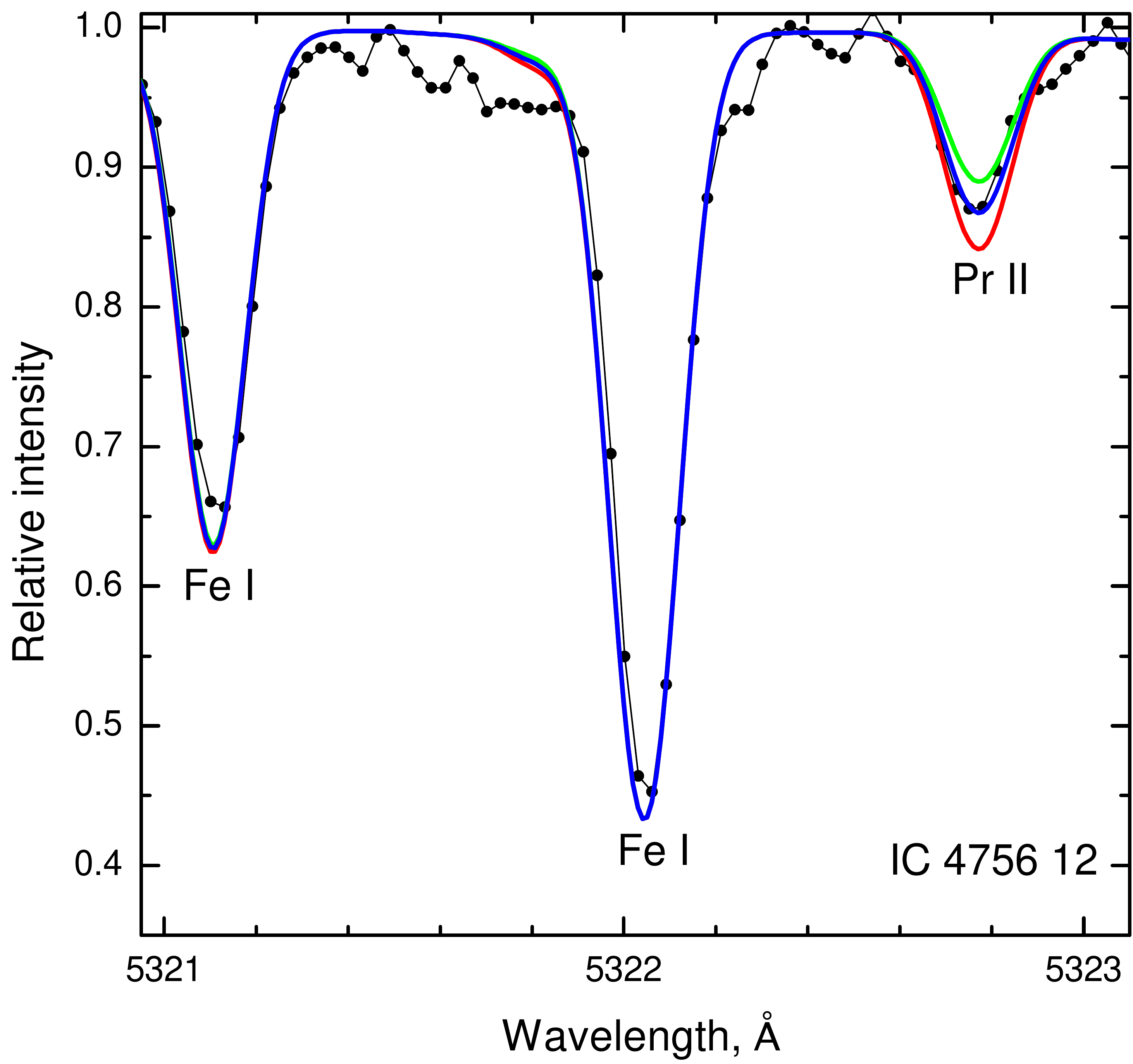}
\includegraphics[scale=0.3]{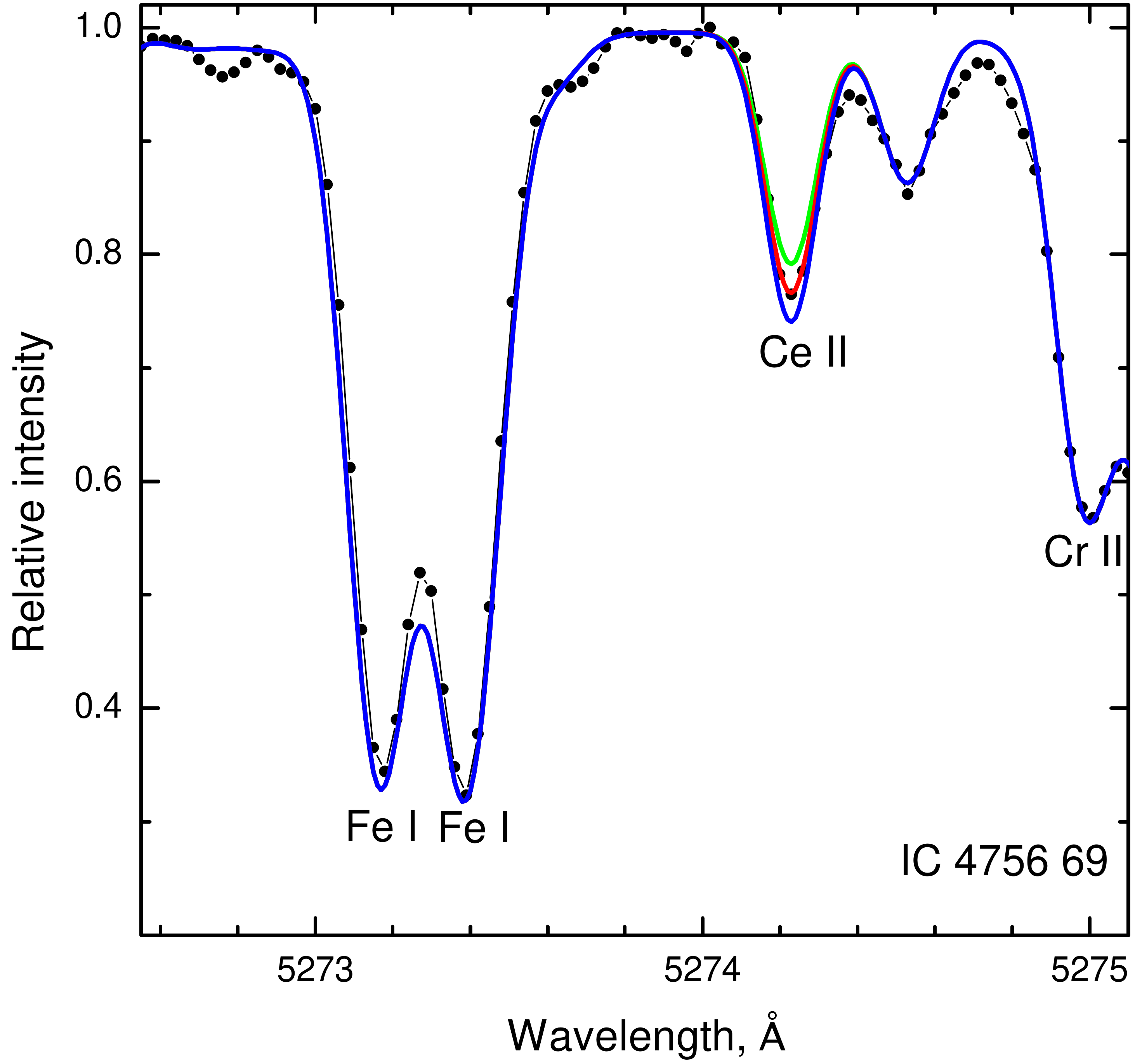}
\includegraphics[scale=0.3]{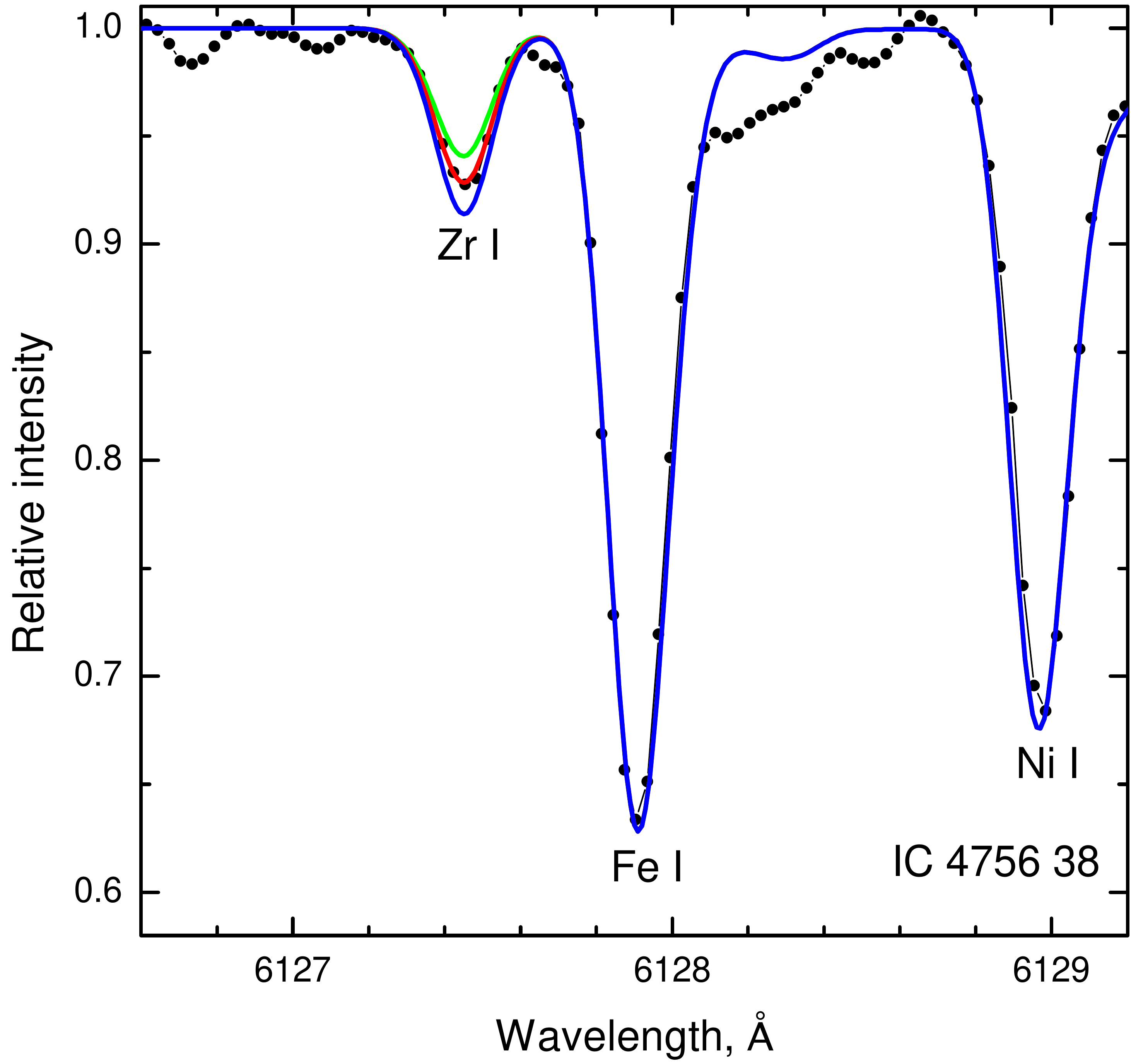}
}
\caption{Examples of the synthetic spectrum fits to various lines for the stars IC\,47561\,12, IC\,47563\,8 and IC\,4756\,69. The blue and green lines represent a change in abundance by $\pm 0.1$~dex to the corresponding elements, except in the case of $^{12}{\rm C}/^{13}{\rm C}$ where the green and red lines represent $\pm 5$.
}
\label{fig:synth_exam}
\end{figure*}
            
\subsection{Membership}
\label{Membership}

The red giants analysed here were selected for observations based on the results of the radial velocity monitoring program of \citet{Mermilliod08}. All the 13 giants in our sample were considered to be likely cluster members. Moreover, all were found to be single stars, apart of star 69 which is a single lined spectroscopic binary. \citet{Mermilliod08} found a mean radial velocity of $-25.16\pm 0.25$~(error) $\pm1.10$~(rms) km\,s$^{-1}$ for IC\,4756. Recently, \citet{Casamiquela16} reported a mean radial velocity of $-$24.7 $\pm$ 0.7~km s$^{-1}$ for the cluster based on the analysis of 7 red giants. The radial velocities determined in our work are within 2~$\sigma$~(rms) of these two mean values for all 12 single giants of the sample. Therefore, as far as radial velocities are concerned, there is no reason to doubt the membership of any of the giants that were observed.

Star 69 is a system with a long period of 1994 days, and a very circular orbit, with eccentricity of 0.05 \citep{Mermilliod07,Swaelmen17}. It was included in the sample analysed here to follow up on the results of \citet{Smiljanic09}, who found star 69 to have very different C and N abundances with respect to other cluster giants (in fact, only limits were determined: [C/Fe] $\leq-0.60$ and [N/Fe] $\geq+0.55$). As suggested in \citet{Smiljanic09}, and further discussed in \citet{Swaelmen17}, star 69 is likely a post-mass-transfer system.

Membership probabilities based on proper motions from UCAC4 \citep{Zacharias13} have been determined for stars in IC\,4756 by \citet{Dias14} and \citet{Sampedro17}. Stars 69, 81, 101, 109 and 125 are part of the \citet{Dias14} catalogue and all five have membership probability of 99\%. Only stars 81 and 109 are part of the \citet{Sampedro17} catalogue. The two have very high membership probabilities ($> 95$\%) according to all three methods used in that work. 

Six of our sample giants were included in the work of \citet{Frinchaboy08}. These authors determined membership probabilities using Tycho-2 proper motions, radial velocities and considering the spatial distribution of the stars. Four stars were considered to be members of the cluster (stars 69, 81, 101, and 109) while stars 44 and 52 (TYC\,455-00950-1 and TYC\,455-00136-1) were found to be non-members (with membership probabilities of 13\% and 47\%, respectively).

Two of our stars (numbers 14 and 109) were included in a sample analysed by \citet{Baumgardt00}. These authors determined membership probabilities for stars in open clusters using Hipparcos parallaxes and proper motions combined with ground-based data (photometry, radial velocity, proper motion, distance from the cluster centre). The membership probabilities determined by \citet{Baumgardt00} for stars 14 and 109 were 59\% and 80\%, respectively.

Another work determining membership probabilities for stars in IC\,4756 was conducted by \citet{Herzog75} using proper motions. Ten of our sample giants were included in that work (all except stars 12, 14, and 42). Only stars 28 and 52 were found to be non-members with membership probabilities of 0\% and 29\%, respectively. For star 44, \citet{Herzog75} found a membership probability of 96\%, which disagrees with the 13\% probability found by \citet{Frinchaboy08} using the Tycho-2 data; the latter probably being more robust. For star 14, the only available study \citep{Baumgardt00} indicates a somewhat low membership probability. Stars 12 and 42 were not included in any of the membership works that used proper motions mentioned here.

Summarizing, there seems to be some evidence that stars 14, 28, 44, and 52 might be non-members of the cluster. In this work, we check how their chemical composition agrees with composition of the high-probability members of the cluster.

\subsection{Atmospheric parameters and elemental abundances}

\begin{table*}
\centering
\caption{A part of table of the spectral line equivalent widths in stars of IC\,4756.}
\label{ew_table}
    \begin{tabular}{cccccccccccccccccccccccccccccccccccccc}
    \hline
    \hline
        \noalign{\smallskip}
Elem.   &       Wavelength      &       \multicolumn{13}{c}{EW (m\AA)}  \\
\cline{3-15}
    \noalign{\smallskip}
& (\AA)         &       12      &       14      &       28      &       38      &       42      &       44      &       52      &       69      &       81      &       101     &       109     &       125     &       164     \\
\hline
\hline
    \noalign{\smallskip}
Na I    &       5148.84 &       33      &       51      &       42      &       --      &       32      &       31      &       52      &       40      &       30      &       35      &       39      &       28      &       33      \\
        &       6154.22 &       67      &       91      &       93      &       67      &       71      &       68      &       107     &       78      &       65      &       68      &       80      &       69      &       74      \\
        &       6160.75 &       89      &       110     &       109     &       85      &       87      &       85      &       123     &       93      &       81      &       84      &       94      &       89      &       94      \\
Mg I    &       6318.70 &       55      &       67      &       67      &       56      &       55      &       53      &       70      &       55      &       46      &       54      &       57      &       55      &       61      \\
...     &       ...     &       ...     &       ...     &       ...     &       ...     &       ...     &       ...     &       ...     &       ...     &       ...     &       ...     &       ...     &       ...     &       ...     \\
        \hline
    \end{tabular}
    \tablefoot{The full table is available at the CDS, http://cdsarc.u-strasbg.fr/}
\end{table*}

We used a standard spectroscopic method, differential to the Sun, for the determination of atmospheric parameters. Effective temperatures ($T_{\rm eff}$) were derived by minimizing a slope between abundances of Fe\,{\sc i} lines with different lower-level excitation potentials ($\chi$). The step of our T$_{\rm eff}$ determination was 5~K, which corresponds to the slope change d\,[Fe/H]/d\,EP = 0.001. Surface gravities (log $\textit{g}$) were determined using the iron ionization equilibrium. Since our step in log\,$g$ determination was 0.1, we allowed  abundances of Fe\,{\sc i} and Fe\,{\sc ii} to differ by no more than 0.02~dex. The final iron abundance was based on the neutral iron lines. In order to find microturbulence velocities ($\textit{v}_{\rm t}$), a minimization of scatter in abundances from Fe\,{\sc i} lines was employed, as well as minimization of the iron abundance trend with regard to the Fe\,{\sc i} line equivalent widths (EWs). Equivalent widths of about 35--38 Fe\, {\sc i} and 4--5 Fe\,{\sc ii} lines were used for the derivation of stellar metallicities as well as other atmospheric parameters. For the measurement of EWs, we used a SPLAT-VO program package \citep{Skoda14}. 

 The EQWIDTH and BSYN software packages (developed at the Uppsala Observatory) were used to derive elemental abundances from EWs and synthetic spectra, respectively.  We have taken a set of plane-parallel, one-dimensional, hydrostatic, constant flux local thermodynamical equilibrium (LTE) model atmospheres \citep{Gustafsson08} from the MARCS stellar model atmosphere and flux library (http://marcs.astro.uu.se/). 
 Atomic oscillator strengths for the main lines used in this work were taken from the inverse solar spectrum analysis performed by \citet{Gurtovenko89}. Using the $gf$ values and solar EWs from \citet{Gurtovenko89}, we computed
the solar elemental abundances used for the differential determination of elemental abundances in the programme stars. For the Sun, we used the main atmospheric parameters 
$T_{\rm{eff}}=5777$~K, log\,$g=4.44$, and the microturbulent velocity value 0.8\,km\,s$^{-1}$, 
as in our earlier studies (e.g. \citealt{Tautvaisiene00}) where the same method of analysis was applied.

Abundances of Na, Mg, Al, Si, Ca, Sc, Ti, Cr, and Ni were determined using the EWs method. The EWs of spectral lines are presented in an online table (see Table~\ref{ew_table} for an example). The number of lines for each element slightly varied among the stars, as every line was inspected individually and some of them were excluded due to contamination of cosmic rays or other observational effects. For sodium, we applied non-local thermodynamic equilibrium (NLTE) corrections as described by \cite{Lind11}. The corrections range from $-0.08$ to $-0.11$~dex.

The spectral synthesis method was used to derive C, N, and O as well as neutron-capture element abundances.
We used the forbidden line at 6300.3~\AA~ for the oxygen abundance determination. The \textit{gf} values for $^{58}$Ni and $^{60}$Ni isotopic line components, which blend the oxygen line, were taken from \cite{Johansson03}. Two C$_2$ molecular bands at 5135~\AA~ and 5635~\AA~ were used to determine carbon abundances and up to eight $^{12}$CN molecular lines in the region of 7980 -- 8005~\AA~ for the nitrogen abundances. We used the same molecular data of $\rm{C}_2$ as in \cite{Gonzalez98} and the CN molecular data provided by Bertrand Plez. The Vienna Atomic Line Data Base (VALD, \citealt{Piskunov95}) was used for preparation of input data used in the calculations. In order to check correctness of the input data, synthetic spectra of the Sun were compared to the solar atlas of \citet{Kurucz05} with solar abundances of  \citep{Grevesse00} and necessary adjustments were made to the line atomic data.

Since carbon and oxygen are bound together by the molecular equilibrium, 
in order to correctly measure abundances of these elements, we investigated them in unison. How abundance changes in one of the elements affect abundances in the others is shown in Table~\ref{CNO_erros}. As we can see, the most sensitive case is between nitrogen and carbon, where 0.1~dex change in carbon coincides with a change in the nitrogen abundance by the same amount. We adopt a procedure of a few back and forth iterations between these elements to achieve a combination of these three abundances, until all of them match the features in the observed spectra. 

\begin{table}
\centering
\caption{Effects on derived abundances and isotopic ratios for the target star IC\,4756\,12, resulting from abundance changes of C, N, or O.}
\label{CNO_erros}
    \begin{tabular}{lrrcccc}
    \\
    \hline
    \hline
    \noalign{\smallskip}
Species & $\Delta$ C & $\Delta$ N & $\Delta$ O \\
 & $\pm0.1$~dex & $\pm0.1$~dex &$\pm0.1$~dex \\
             \hline
   \hline
                   \noalign{\smallskip}
$\Delta$ C                      &       --              &        $\pm$0.01      &        $\pm$0.02       \\
$\Delta$ N                      &       $\mp$0.10       &       --              &        $\pm$0.05       \\
$\Delta$ O                      &       $\pm$0.01       &        $\pm$0.01      &       --              \\
$\Delta$ C/N            &       $\pm$0.19       &        $\mp$0.16      &        $\mp$0.01       \\
$\Delta ^{12}{\rm C}/^{13}{\rm C}$      &        $\pm$2         &        $\pm$2  &        0      \\
    \hline
\end{tabular}
\end{table}

\begin{table}
        \centering
    \caption{Effects on derived abundances, $\Delta$[A/H], resulting from model changes for the star IC\,4756\,12. }
    \label{uncertainties_parameters}
    \begin{tabular}{lrrrrc}
    \hline
    \noalign{\smallskip}
Species & ${ \Delta T_{\rm eff} }\atop{ \pm100~{\rm~K} }$ & ${ \Delta \log g }\atop{ \pm0.3 }$ & ${ \Delta v_{\rm t} }\atop{ \pm0.3~{\rm km~s}^{-1}}$ &
             ${ \Delta {\rm [Fe/H]} }\atop{ \pm0.1}$ & Total \\
    \noalign{\smallskip}
    \hline
    \noalign{\smallskip}
C (C$_{2}$) &   0.03    &       0.02    &       0.00    &       0.01    &       0.04    \\
N (CN) &        0.10            & 0.03  &       0.01    &       0.01    &        0.11  \\
O ({[O\,{\sc i}]}) &    0.01            &       0.14    &       0.00    &       0.01    &       0.14   \\
$^{12}{\rm C}/^{13}{\rm C}$ &   1       &        1      &       0       &       0       &        1.4     \\
\noalign{\smallskip}
Na\,{\sc i}     &       $0.06$  &       $-0.01$ &       $0.00$  &       $0.00$  &       $0.06$  \\
Mg\,{\sc i}             &       $0.05$  &       $0.00$  &       $0.01$  &       $-0.01$ &       $0.05$  \\
Al\,{\sc i}             &       $0.06$  &       $0.00$  &       $0.01$  &       $-0.01$ &       $0.06$  \\
Si\,{\sc i}             &       $0.01$  &       $0.04$  &       $0.00$  &       $0.01$  &       $0.04$  \\
Ca\,{\sc i}             &       $0.08$  &       $-0.01$ &       $0.01$  &       $-0.01$ &       $0.08$  \\
Sc\,{\sc ii}            &       $-0.01$ &       $0.13$  &       $0.00$  &       $0.03$  &       $0.13$  \\
Ti\,{\sc i}             &       $0.11$  &       $-0.01$ &       $0.00$  &       $-0.01$ &       $0.11$  \\
Ti\,{\sc ii}            &       $-0.01$ &       $0.13$  &       $-0.01$ &       $0.03$  &       $0.13$  \\
Fe\,{\sc i}             & $ 0.09$               &       $0.01$  &       $0.00$ & $0.01$                & $0.09$                \\
Fe\,{\sc ii}            &       $-0.07$ &       $0.14$  &       $-0.01$ &       $0.04$  &       $0.15$  \\
Cr\,{\sc i}             &       $0.10$  &       $-0.01$ &       $-0.01$ &       $-0.01$ &       $0.10$  \\
Mn\,{\sc i}             &       $0.12$  &       $0.04$  &       $-0.11$ &       $0.01$  &       $0.17$  \\
Co\,{\sc i}             &       $0.10$  &       $0.03$  &       $0.02$  &       $0.01$  &       $0.11$  \\
Ni\,{\sc i}             &       $0.06$  &       $0.03$  &       $0.00$  &       $0.01$  &       $0.07$  \\
Y\,{\sc ii}             &       $0.01$  &       $0.11$  &       $-0.25$ &       $0.04$  &       $0.27$  \\
Zr\,{\sc i}             &       $0.15$  &       $0.00$  &       $0.02$  &       $0.00$  &       $0.15$  \\
Ba\,{\sc ii}            &       $0.05$  &       $0.11$  &       $-0.36$ &       $0.01$  &       $0.38$  \\
La\,{\sc ii}            &       $0.02$  &       $0.13$  &       $-0.01$ &       $0.03$  &       $0.13$  \\
Ce\,{\sc ii}            &       $0.01$  &       $0.13$  &       $-0.05$ &       $0.04$  &       $0.14$  \\
Pr\,{\sc ii}            &       $0.01$  &       $0.13$  &       $-0.01$ &       $0.03$  &       $0.13$  \\
Nd\,{\sc ii}            &       $0.03$  &       $0.14$  &       $-0.05$ &       $0.03$  &       $0.15$  \\
Eu\,{\sc ii}            &       $-0.01$ &       $0.11$  &       $0.00$  &       $0.03$  &       $0.11$  \\
    \noalign{\smallskip}
    \hline
    \end{tabular}
\end{table}

\begin{table*}
        \centering
    \caption{Atmospheric parameters of the programme stars, the Sun, and Arcturus.}
    \label{stellar_parameters}
    \begin{tabular}{lccccrcccc}
    \hline
    \hline
    \noalign{\smallskip}
Star ID &       $T_{\rm eff}$ (K)       &       log \textit{g}  &       $v_{\rm t}$ (${\rm km~s}^{-1}$) &       $v$\,sin\,\textit{i} (${\rm km~s}^{-1}$)        &       {[Fe/H]}        &       $ \sigma $\,Fe\,{\sc i}   &       n       &  $ \sigma $\,Fe\,{\sc ii}     &       n       \\
        \hline
        \hline
        \noalign{\smallskip}
        \multicolumn{10}{c}{Members}\\
12      &       5135    &       2.6     &       1.45    &       1.6     &       $-0.03$ &       0.06    &       38      &       0.02    &       4       \\
38      &       5165    &       2.9     &       1.35    &       1.9     &       0.00    &       0.07    &       38      &         0.05    &       5       \\
42      &       5165    &       2.7     &       1.25    &       3.0     &       $-0.01$ &       0.04    &       36      &         0.06    &       5       \\
69      &       5150    &       2.7     &       1.40    &       2.5     &       0.00    &       0.07    &       38      &         0.04    &       5       \\
81      &       5180    &       2.8     &       1.15    &       3.7     &       $-0.01$ &       0.06    &       37      &         0.06    &       5       \\
101     &       5135    &       2.8     &       1.40    &       3.2     &       $-0.01$ &       0.07    &       38      &         0.02    &       5       \\
109     &       5000    &       2.6     &       1.50    &       2.9     &       $-0.03$ &       0.06    &       38      &         0.04    &       5       \\
125     &       5150    &       2.8     &       1.45    &       2.6     &       $-0.02$ &       0.05    &       38      &         0.02    &       5       \\
164     &       5040    &       2.6     &       1.45    &       2.2     &       $-0.03$ &       0.06    &       38      &         0.04    &       5       \\
    \noalign{\smallskip}
        \multicolumn{10}{c}{Doubtful members}\\
14      &       4760    &       2.3     &       1.50    &       1.9     &       $-0.06$ &       0.07    &       37      &         0.04    &       4       \\
28      &       4650    &       2.1     &       1.45    &       2.2     &       $-0.10$ &       0.08    &       38      &         0.06    &       5       \\
44      &       5115    &       2.7     &       1.25    &       3.6     &       $-0.02$ &       0.06    &       37      &         0.07    &       5       \\
52      &       4500    &       1.9     &       1.60    &       2.8     &       $-0.12$ &       0.08    &       35      &         0.05    &       5       \\
    \noalign{\smallskip}
        \noalign{\smallskip}
Sun &    5777   &    4.44   &     0.8   &    2   &      0.05$\text{*}$      &    0.03      &   38    &     0.01      &   5   \\
Arcturus & 4345 & 1.6 & 1.65 & 2.4 & $-0.58$ & 0.04 & 38 & 0.03 & 5 \\
    \hline
    \end{tabular}
    \tablefoot{$\text{*}$ The solar iron abundance in this work is A(Fe)$_{\odot}$=7.50 (\citealt{Grevesse00}).}
\end{table*}

The synthetic spectra method was also used for the determination of (Mn, Co, Y, Zr, Ba, La, Ce, Pr, Nd, and Eu abundances. 
Cobalt abundances were determined from the 5280.62, 5301.03, 5352.05, 5530.78, 5647.23, 6188.98, 6455, and 6814.95~\AA~ lines. For the analysis of lines at 5301.03 and 5530.78\,\AA~ we applied hyperfine structure (HFS) data from \cite{Nitz99}, while for the remaining Co\,{\sc i} lines the HFS data were taken from \cite{Cardon82}. Manganese was investigated using lines at 6013.49, 6016.64, and 6021.80~\AA~ with the HFS data taken from \cite{Hartog11}.    
Yttrium abundances were determined from the Y\,{\sc ii} lines at 4883.69, 4900.12, 4982.14, 5200.41, and 5402.78~\AA; zirconium from the Zr\,{\sc i} lines at 5385.10 and 6127.50~\AA; lanthanum from the La\,{\sc ii} lines at 5123.01, 6320.41, and 6390.48~\AA. For the analysis of the La\,{\sc ii}  5123.01~\AA~ and 6390.48~\AA~ lines, we applied the HFS data from \citep{Ivans06}. We were not able to find the HFS data for the La\,{\sc ii} 6320.41~\AA~ line, however it seems that the HFS influence is small for this line since lanthanum abundances were very similar from all three lines. Cerium abundances were determined from the Ce\,{\sc ii} lines at 5274.22, 6043.00~\AA. Neodymium abundances were derived from the Nd\,{\sc ii} lines at 5092.80, 5293.20, 5319.80, 5356.97 and 5740.86~\AA~ with the HFS adopted from \citet{DenHartog03}. 
Barium and europium abundances were determined from single lines at 5853.67~\AA~ and 6645.10~\AA, respectively. For the Ba\,{\sc ii} line, the HFS data were taken from \citet{McWilliam98} and for the Eu\,{\sc ii} line from \citet{Lawler01}. The HFS was also taken into account for the determination of praseodymium abundance from Pr\,{\sc ii}  lines at 5259.7~\AA~ and 5322.8~\AA~ \citep{Sneden09}. All log\,$gf$ values were calibrated to fit the solar spectrum by \citet{Kurucz05} with solar abundances provided by  \citep{Grevesse00}.
Several examples of the synthetic spectra fits for some of the lines are presented in Fig.~\ref{fig:synth_exam}. 

As stellar rotation ($v\,{\rm sin}\,i$) changes the shape of lines, it is also an important factor in abundance determinations from spectral syntheses. We obtained $v\,{\rm sin}\,i$ by fitting iron lines of different strengths in our investigated spectral regions. 

There are two categories of uncertainties that should be considered. Firstly, there are random errors that affect each line independently and originate from the local continuum placement, from each line's fitting variations, EW measurements, or from uncertainties in atomic parameters. Uncertainties coming from the atomic data are minimized when a differential analysis is performed.
Since all cluster-member stars have relatively similar parameters in our case, we can compare measurements of EWs for the same line in different stars. The mean scatter of EWs for iron lines was 4.3 m\AA. 
An approximate value for the random errors in the abundance determinations is given by the scatter of the derived abundances from individual lines for all elements in all stars. The mean scatter in our sample of stars is 0.06~dex. 

We evaluated uncertainties of atmospheric parameters using the whole sample of IC\,4756 member stars. The average of slopes between [Fe/H] abundance and the excitation potential (EP) for these stars d\,[Fe/H]/d\,EP = $0.0008 \pm 0.008$. The error of 0.008 corresponds to the temperature change of 30~K. We can assume this as an error in our temperature estimation. The change of log\,$g$ by 0.1~dex alters the Fe\,{\sc i} and Fe\,{\sc ii} equilibrium by 0.04~dex, which is larger than our acceptable tolerance of 0.02~dex. Therefore, the error in log\,$g$ determination is less than 0.1~dex. For the microturbulence velocity we minimized a slope between the Fe\,{\sc i} abundances and the EWs. The mean  d\,[Fe/H]/d\,EW =$ -0.0001 \pm 0.0003$. This error corresponds to the change of 0.05~km\,s$^{-1}$, which can be assumed as our error for the microtubulence velocity determination. 

The second type of uncertainties are systematic; they are influenced by uncertainties of atmospheric parameters and affect all the lines simultaneously. Table~\ref{uncertainties_parameters} shows sensitivity of abundance estimates to changes in atmospheric parameters for the star IC\,4756\,12. Changes in atmospheric parameters provide relatively small abundance deviations from the initial values. The larger deviations are present for abundances derived from ionized lines, as these elements respond to the log\,$g$ value changes more strongly.
 
Along with IC\,4756 stars, we performed our atmospheric parameter and abundance determination procedures on the Arcturus's spectrum by \citep{Hinkle00} (the results for Arcturus we present together with the IC\,4756 results).
Our results for Arcturus agree well with the majority of recent studies of this star (\citealt{Jofre15, Abia12, Ramirez11, Worley09}).

Several examples of the synthetic spectra fits in several IC\,4756 stars for some of the lines are presented in Fig.~\ref{fig:synth_exam}.

\begin{table*}                                                                                  
        \centering                                                                                      
    \caption{Elemental abundances in member stars of IC\,4756 and Arcturus.}                                                                                    
    \label{all_star_abundances}                                                                                 
    \begin{tabular}{lccccc}                                                                                     
    \hline                                                                                      
    \hline                                                                                      
    \noalign{\smallskip}                                                                                        
El./Star                &               12              &               38              &               42              &               69              &               81              \\
    \hline                                                                                      
            \noalign{\smallskip}                                                                                        
{[C/Fe]}                &                $-0.23\pm0.01$\,(02)           &               $-0.23\pm0.03$\,(02)            &                $-0.29\pm0.02$\,(02)            &               --              &               $-0.29\pm0.02$\,(02)            \\
{[N/Fe]}                &                $0.45\pm0.06$\,(08)            &               $0.43\pm0.03$\,(08)             &               $0.45\pm0.04$\,(08)             &               --              &               $0.42\pm0.05$\,(07)             \\
{[O/Fe]}                &                0.01           &               0.09            &                $-0.10$         &               --              &               $-0.05$         \\
C/N             &               0.83            &               0.87            &               0.72            &               --              &               0.78            \\
$^{12}{\rm C}/^{13}{\rm C}$             &               19              &               20              &               --              &               --              &               17              \\
$\rm [Na/Fe]_{LTE}$             &               $0.23\pm0.04$\,(03)             &               $0.17\pm0.01$\,(02)             &               $0.25\pm0.02$\,(03)             &               $0.33\pm0.04$\,(03)             &               $0.19\pm0.02$\,(03)             \\
$\rm [Na/Fe]_{NLTE}$            &        $0.15\pm0.07$\,(03)            &               $0.06\pm0.02$\,(02)             &               $0.17\pm0.04$\,(03)             &       $0.24\pm0.08$\,(03)             &               $0.10\pm0.06$\,(03)             \\
{[Mg/Fe]}               &               $0.08\pm0.04$\,(02)             &               $0.11\pm0.01$\,(02)             &               $0.15\pm0.05$\,(02)             &               $0.05\pm0.04$\,(02)             &               $-0.02$\,(01)           \\
{[Al/Fe]}               &               $0.10\pm0.01$\,(02)             &               $0.00\pm0.02$\,(02)             &               $0.01\pm0.08$\,(02)             &               $0.00\pm0.07$\,(02)             &               $0.03\pm0.08$\,(02)             \\
{[Si/Fe]}               &               $0.00\pm0.07$\,(16)             &               $0.05\pm0.08$\,(19)             &               $0.05\pm0.06$\,(19)             &               $0.01\pm0.06$\,(17)             &               $-0.01\pm0.05$\,(18)            \\
{[Ca/Fe]}               &               $0.06\pm0.09$\,(06)             &               $0.13\pm0.08$\,(08)             &               $0.11\pm0.10$\,(07)             &               $0.10\pm0.08$\,(07)             &               $0.10\pm0.06$\,(08)             \\
{[Sc\,{\sc ii}/Fe]}             &               $-0.08\pm0.05$\,(07)            &               $0.03\pm0.02$\,(07)             &               $-0.06\pm0.05$\,(07)            &               $-0.02\pm0.05$\,(07)            &               $0.00\pm0.06$\,(06)             \\
{[Ti\,{\sc i}/Fe]}              &               $-0.02\pm0.04$\,(13)            &               $0.00\pm0.04$\,(16)             &               $-0.02\pm0.04$\,(14)            &               $-0.01\pm0.04$\,(15)            &               $0.01\pm0.05$\,(14)             \\
{[Ti\,{\sc ii}/Fe]}             &               $-0.11\pm0.07$\,(05)            &               $-0.01\pm0.02$\,(05)            &               $-0.10\pm0.05$\,(05)            &               $-0.04\pm0.07$\,(05)            &               $-0.08\pm0.04$\,(05)            \\
{[Cr/Fe]}               &               $0.08\pm0.05$\,(08)             &               $0.02\pm0.07$\,(09)             &               $0.03\pm0.06$\,(09)             &               $0.05\pm0.08$\,(10)             &               $0.01\pm0.07$\,(10)             \\
{[Mn/Fe]}               &               $-0.08\pm0.02$\,(03)            &               $-0.02\pm0.01$\,(03)            &               $-0.09\pm0.05$\,(03)            &               $-0.07\pm0.06$\,(03)            &               $-0.19\pm0.02$\,(03)            \\
{[Co/Fe]}               &               $-0.02\pm0.03$\,(08)            &               $-0.03\pm0.03$\,(08)            &               $-0.07\pm0.02$\,(08)            &               $-0.03\pm0.03$\,(08)            &               $-0.07\pm0.02$\,(08)            \\
{[Ni/Fe]}               &               $-0.08\pm0.09$\,(24)            &               $-0.06\pm0.07$\,(26)            &               $-0.06\pm0.10$\,(25)            &               $-0.06\pm0.08$\,(26)            &               $-0.07\pm0.07$\,(21)            \\
{[Y/Fe]}                &               $-0.05\pm0.14$\,(05)            &               $0.08\pm0.09$\,(05)             &               $0.02\pm0.05$\,(05)             &               $0.06\pm0.08$\,(05)             &               $-0.01\pm0.08$\,(05)            \\
{[Zr/Fe]}               &               $0.20$\,(01)            &               $0.24\pm0.14$\,(02)             &               $0.17\pm0.02$\,(02)             &               $0.14\pm0.04$\,(02)             &               $0.18$\,(01)            \\
{[Ba/Fe]}               &               $0.20$\,(01)            &               $0.23$\,(1)             &               $0.24$\,(1)             &               $0.30$\,(01)            &               $0.22$\,(01)            \\
{[La/Fe]}               &               $0.20\pm0.02$\,(03)             &               $0.23\pm0.05$\,(03)             &               $0.14\pm0.02$\,(03)             &               $0.19\pm0.03$\,(02)             &               $0.23\pm0.04$\,(03)             \\
{[Ce/Fe]}               &               $0.21\pm0.10$\,(02)             &               $0.27\pm0.11$\,(02)             &               $0.18\pm0.07$\,(02)             &               $0.17\pm0.06$\,(02)             &               $0.18\pm0.04$\,(02)             \\
{[Pr/Fe]}               &               $0.21\pm0.03$\,(02)             &               $0.18\pm0.07$\,(02)             &               $0.10\pm0.01$\,(02)             &               $0.14\pm0.06$\,(02)             &               $0.16\pm0.02$\,(02)             \\
{[Nd/Fe]}               &               $0.08\pm0.06$\,(05)             &               $0.14\pm0.06$\,(05)             &               $0.03\pm0.06$\,(05)             &               $0.13\pm0.04$\,(05)             &               $0.11\pm0.07$\,(05)             \\
{[Eu/Fe]}               &               $0.02$\,(01)            &               $0.11$\,(01)            &               $0.03$\,(01)            &               $0.04$\,(01)            &               $0.03$\,(01)            \\
    \noalign{\smallskip}                                                                                        
        \cline{1-6}                                                                                     
    \cline{1-6}                                                                                 
        \noalign{\smallskip}                                                                                    
El./Star                &               101             &               109             &               125             &               164      & Arcturus      \\                      
\cline{1-6}                                                                                     
        \noalign{\smallskip}                                                                                    
{[C/Fe]}                &                $-0.23\pm0.03$\,(02)           &               $-0.24\pm0.05$\,(02)            &               $-0.26$\,(01)           &               $-0.28\pm0.02$\,(02) &       $0.03\pm0.01$\,(02)\\                   
{[N/Fe]}                &               $0.46\pm0.04$\,(07)             &               $0.47\pm0.03$\,(08)             &               $0.46\pm0.05$\,(08)             &               $0.45\pm0.03$\,(08)      & $0.29\pm0.02$\,(08)\\                         
{[O/Fe]}                &               0.05            &               0.02            &               0.02            &               0.02 &       0.55 \\                         
C/N             &               0.81            &               0.78            &               0.76            &               0.74 &       2.19 \\                         
$^{12}{\rm C}/^{13}{\rm C}$             &               --              &               --              &               17              &               20      & 7 \\                                            $\rm [Na/Fe]_{LTE}$             &               $0.19\pm0.06$\,(03)             &               $0.24\pm0.04$\,(03)             &               $0.19\pm0.04$\,(03)             &               $0.22\pm0.02$\,(03)      & $0.20\pm0.05$\,(03)\\                         
$\rm [Na/Fe]_{NLTE}$            &               $0.11\pm0.11$\,(03)             &               $0.16\pm0.09$\,(03)             &               $0.11\pm0.01$\,(03)             &               $0.14\pm0.03$\,(03)      & $0.16\pm0.06$\,(03)\\                         
{[Mg/Fe]}               &               $0.03\pm0.05$\,(02)             &               $0.04\pm0.02$\,(02)             &               $0.11\pm0.01$\,(02)             &               $0.14\pm0.02$\,(02)     & $0.43\pm0.01$\,(02)\\                           
{[Al/Fe]}               &               $0.01\pm0.05$\,(02)             &               $0.05\pm0.04$\,(02)             &               $0.07\pm0.04$\,(02)             &               $0.06\pm0.01$\,(02)     & $0.43\pm0.07$\,(03)\\                           
{[Si/Fe]}               &               $0.06\pm0.07$\,(18)             &               $0.08\pm0.08$\,(17)             &               $0.06\pm0.06$\,(18)             &               $0.07\pm0.07$\,(17)     & $0.25\pm0.04$\,(18)\\                           
{[Ca/Fe]}               &               $0.09\pm0.06$\,(07)             &               $0.16\pm0.09$\,(07)             &               $0.12\pm0.08$\,(07)             &               $0.12\pm0.06$\,(08)     & $0.18\pm0.06$\,(08)\\                           
{[Sc\,{\sc ii}/Fe]}             &               $0.01\pm0.06$\,(06)             &               $0.01\pm0.07$\,(07)             &               $0.03\pm0.07$\,(07)             &               $0.01\pm0.06$\,(07)     & $0.15\pm0.04$\,(07)\\                           
{[Ti\,{\sc i}/Fe]}              &               $-0.04\pm0.04$\,(13)            &               $0.01\pm0.05$\,(12)             &               $0.02\pm0.05$\,(15)             &               $0.01\pm0.04$\,(13)     & $0.27\pm0.05$\,(16)\\                           
{[Ti\,{\sc ii}/Fe]}             &               $0.02\pm0.06$\,(05)             &               $0\pm0.05$\,(05)                &               $-0.02\pm0.02$\,(05)            &               $-0.04\pm0.04$\,(05) & $0.23\pm0.05$\,(04)   \\                              
{[Cr/Fe]}               &               $-0.01\pm0.04$\,(09)            &               $0.00\pm0.07$\,(09)             &               $0.01\pm0.06$\,(10)             &               $0.04\pm0.06$\,(09)     & $-0.03\pm0.07$\,(10)\\                          
{[Mn/Fe]}               &               $-0.06\pm0.05$\,(03)            &               $-0.07\pm0.06$\,(03)            &               $-0.02\pm0.03$\,(03)            &               $-0.04\pm0.07$\,(03)    & $-0.18\pm0.08$\,(03)\\                          
{[Co/Fe]}               &               $-0.06\pm0.03$\,(08)            &               $-0.01\pm0.03$\,(08)            &               $-0.02\pm0.02$\,(08)            &               $-0.04\pm0.04$\,(08)    & $0.2\pm0.03$\,(08)\\                            
{[Ni/Fe]}               &               $-0.05\pm0.07$\,(25)            &               $-0.05\pm0.08$\,(24)            &               $-0.04\pm0.05$\,(24)            &               $-0.03\pm0.05$\,(26) &       $0.02\pm0.09$\,(24)\\                           
{[Y/Fe]}                &               $0.06\pm0.10$\,(05)             &               $0.03\pm0.10$\,(05)             &               $0.09\pm0.06$\,(05)             &               $0.02\pm0.12$\,(05)     & $-0.08\pm0.16$\,(05)\\                          
{[Zr/Fe]}               &               $0.11\pm0.10$\,(02)             &               $0.22\pm0.08$\,(02)             &               $0.21\pm0.07$\,(02)             &               $0.19\pm0.04$\,(02)     & $-0.02\pm0.11$\,(02)\\                          
{[Ba/Fe]}               &               $0.25$\,(01)            &               $0.18\pm0$\,(1)         &               $0.23$\,(01)            &               $0.18$\,(01) &       $-0.22$\,(01)\\                         
{[La/Fe]}               &               $0.27\pm0.03$\,(03)             &               $0.24\pm0.03$\,(03)             &               $0.26\pm0.01$\,(02)             &               $0.24\pm0.01$\,(03)     & $0.07\pm0.05$\,(03)\\                           
{[Ce/Fe]}               &               $0.21\pm0.06$\,(02)             &               $0.24\pm0.11$\,(02)             &               $0.24\pm0.08$\,(02)             &               $0.21\pm0.07$\,(02)     & $-0.10\pm0.11$\,(02)\\                          
{[Pr/Fe]}               &               $0.19\pm0.03$\,(02)             &               $0.22\pm0.06$\,(02)             &               $0.22\pm0.04$\,(02)             &               $0.19\pm0.01$\,(02)     & $0.19\pm0.04$\,(02)\\                           
{[Nd/Fe]}               &               $0.16\pm0.07$\,(05)             &               $0.16\pm0.03$\,(05)             &               $0.15\pm0.06$\,(05)             &               $0.13\pm0.05$\,(05)     & $-0.08\pm0.04$\,(05)\\                          
{[Eu/Fe]}               &               $0.10$\,(01)            &               $0.08$\,(01)            &               $0.13$\,(01)            &               $0.14$\,(01) &       $0.38$\,(01)\\                          
    \noalign{\smallskip}                                                                                        
\cline{1-6}                                                                                     
    \end{tabular}                                                                                       
        \tablefoot{                                                                                     
   The elemental abundance ratios are presented together with the abundance scatter from individual lines and a number of lines used for the analysis.                                                                                         
    }                                                                                   
\end{table*}

\begin{table*}                                                          
        \centering                                                              
    \caption{Elemental abundances in doubtful member stars of IC\,4756.}                                                                                                                      
    \label{all_star_abundances}                                                         
    \begin{tabular}{lcccc}                                                              
    \hline                                                              
    \hline                                                              
    \noalign{\smallskip}                                                                
El./Star                &               14              &               28              &               44              &               52      \\
    \hline                                                              
            \noalign{\smallskip}                                                                
{[C/Fe]}                &               $-0.24\pm0.03$\,(02)            &               $-0.20$\,(01)           &               $-0.32\pm0.02$\,(02)            &               $-0.18\pm0.03$\,(02)    \\
{[N/Fe]}                &               $0.50\pm0.03$\,(08)             &               $0.37\pm0.02$\,(08)             &               $0.42\pm0.04$\,(08)             &               $0.49\pm0.05$\,(08)     \\
{[O/Fe]}                &               0.10            &               0.09            &               $-0.08$         &               0.14    \\
C/N             &                0.72           &               --              &               0.72            &               0.85    \\
$^{12}{\rm C}/^{13}{\rm C}$             &               19              &               --              &               --              &               --      \\
$\rm [Na/Fe]_{LTE}$             &               $0.31\pm0.03$\,(03)             &               $0.24\pm0.06$\,(03)             &               $0.19\pm0.01$\,(03)             &               $0.28\pm0.06$\,(03)     \\
$\rm [Na/Fe]_{NLTE}$            &               $0.23\pm0.07$\,(03)             &               $0.16\pm0.03$\,(03)             &               $0.10\pm0.05$\,(03)             &               $0.20\pm0.04$\,(03)     \\
{[Mg/Fe]}               &               $0.13\pm0.03$\,(02)             &               $0.14\pm0.01$\,(02)             &               $0.03\pm0.06$\,(02)             &               $0.12\pm0.02$\,(02)     \\
{[Al/Fe]}               &               $0.09\pm0.04$\,(02)             &               $0.14\pm0.04$\,(02)             &               $0.06\pm0.05$\,(02)             &               $0.19\pm0.07$\,(02)     \\
{[Si/Fe]}               &               $0.13\pm0.1$\,(17)              &               $0.16\pm0.10$\,(17)             &               $0.03\pm0.08$\,(16)             &               $0.18\pm0.10$\,(13)     \\
{[Ca/Fe]}               &               $0.14\pm0.10$\,(07)             &               $0.08\pm0.09$\,(05)             &               $0.12\pm0.06$\,(07)             &               $0.17\pm0.11$\,(08)     \\
{[Sc\,{\sc ii}/Fe]}             &               $0.08\pm0.04$\,(06)             &               $0.02\pm0.09$\,(06)             &               $0.01\pm0.03$\,(05)             &               $0.04\pm0.10$\,(06)     \\
{[Ti\,{\sc i}/Fe]}              &               $0.02\pm0.05$\,(16)             &               $0.02\pm0.07$\,(17)             &               $0.01\pm0.06$\,(15)             &               $0.00\pm0.08$\,(18)     \\
{[Ti\,{\sc ii}/Fe]}             &               $0.03\pm0.02$\,(05)             &               $-0.02\pm0.02$\,(05)            &               $-0.06\pm0.05$\,(05)            &               $0.07\pm0.03$\,(05)     \\
{[Cr/Fe]}               &               $0.02\pm0.08$\,(10)             &               $0.03\pm0.07$\,(09)             &               $0.07\pm0.10$\,(10)             &               $0.02\pm0.07$\,(09)     \\
{[Mn/Fe]}               &               $-0.12\pm0.05$\,(03)            &               $-0.07\pm0.08$\,(03)            &               $-0.16\pm0.02$\,(03)            &               $-0.07\pm0.08$\,(03)    \\
{[Co/Fe]}               &               $-0.01\pm0.02$\,(08)            &               $-0.02\pm0.04$\,(08)            &               $-0.07\pm0.04$\,(08)            &               $0.02\pm0.04$\,(08)     \\
{[Ni/Fe]}               &               $0.00\pm0.11$\,(21)             &               $-0.02\pm0.11$\,(24)            &               $-0.07\pm0.07$\,(26)            &               $0.03\pm0.12$\,(24)     \\                                                                                                                                      
{[Y/Fe]}                &               $0.01\pm0.10$\,(05)             &               $0.05\pm0.12$\,(05)             &               $0.01\pm0.07$\,(05)             &               $0.05\pm0.15$\,(05)     \\
{[Zr/Fe]}               &               $0.21\pm0.06$\,(02)             &               $0.15\pm0.01$\,(02)             &               $0.16\pm0.09$\,(02)             &               $0.13\pm0.02$\,(02)     \\
{[Ba/Fe]}               &               $0.15$\,(01)            &               $0.19$\,(01)            &               $0.27$\,(01)            &               $0.19$\,(01)    \\
{[La/Fe]}               &               $0.30\pm0.06$\,(03)             &               $0.23\pm0.05$\,(03)             &               $0.20\pm0.07$\,(03)             &               $0.30\pm0.06$\,(03)     \\
{[Ce/Fe]}               &               $0.24\pm0.14$\,(02)             &               $0.21\pm0.13$\,(02)             &               $0.20\pm0.15$\,(02)             &               $0.29\pm0.12$\,(02)     \\
{[Pr/Fe]}               &               $0.27\pm0.04$\,(02)             &               $0.22\pm0.01$\,(02)             &               $0.14\pm0.03$\,(02)             &               $0.31\pm0.06$\,(02)     \\
{[Nd/Fe]}               &               $0.10\pm0.04$\,(05)             &               $0.09\pm0.04$\,(05)             &               $0.04\pm0.05$\,(05)             &               $0.15\pm0.08$\,(05)     \\
{[Eu/Fe]}               &               $0.15$\,(01)            &               $0.14$\,(01)            &               $0.04$\,(01)            &               $0.17$\,(01)    \\
    \noalign{\smallskip}
    \hline
    \end{tabular}                                                               
        \tablefoot{                                                             
   The elemental abundance ratios are presented together with the abundance scatter from individual lines and a number of lines used for the analysis.                                                                 
    }                                                           
\end{table*}                                                            

\section{Results and discussion}

The determined stellar atmospheric parameters and rotational velocities for our IC\,4756 programme stars are  presented in Table~\ref{stellar_parameters}, and the detailed chemical abundances are in Table~\ref{all_star_abundances}. Mean values of element-to-iron ratios were calculated taking just definite members of IC\,4756. These values with corresponding scatters are listed in Table~\ref{authorcomparison} along with results from  previous studies. \citet{Luck94} investigated one supergiant, one dwarf, and two giants. Since the investigated dwarf star was rather peculiar, we did not take its abundances while calculating the mean values presented in Table~\ref{authorcomparison}.  

\subsection{Atmospheric parameters}

 The average metallicity for this cluster determined in this work from nine high-probability members is close to solar – [Fe/H] = $ -0.02\pm 0.01$. Those stars have similar atmospheric parameters and are in the red clump: the average $T_{\rm eff} = 5124 \pm 58$~K, log\,$\textit{g} = 2.72 \pm 0.1$, [Fe/H]$ = -0.02 \pm 0.01$, $v_{\rm t} = 1.38 \pm 0.11$~km\,s$^{-1}$. 
 
 Our mean metallicity value agrees well with the majority of the previous spectroscopic studies (\citealt{Gilroy89}, \citet{Luck94}, \citet{Smiljanic09}, \citet{Santos09}, \citet{Ting12}), all of which provide metallicities close to solar for this cluster. Several studies by \citet{Thogersen93} and \citet{Jacobson07} derived a slightly sub-solar metallicity of $-0.22\pm0.12$~dex and $-0.15\pm0.04$~dex, respectively. These discrepant metallicities were probably caused by different line lists and/or analysis techniques.

 The stars 14, 28, 44, and 52, as already discussed in Sect.~\ref{Membership}, are considered as doubtful members according to the literature, however we chose to investigate them and compare their chemical composition to other stars of the cluster. We found that stars 14, 28, and 52 have slightly lower metallicities thus we support their doubtful membership. Star 44 has the same metallicity as the cluster average, thus we leave its membership question open.

\begin{figure}
   \centering
        \includegraphics[width=\columnwidth]{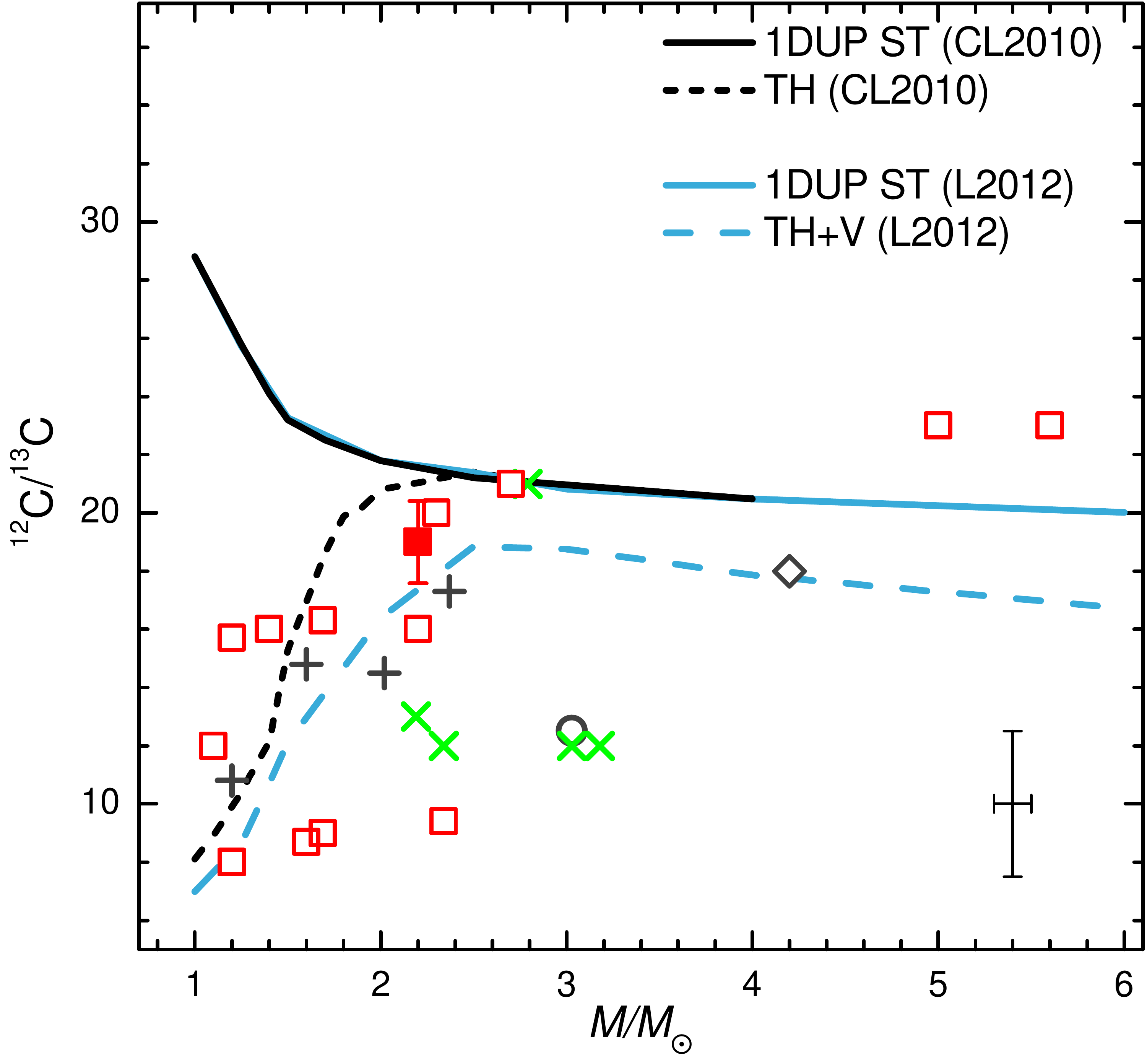}
   \caption{The average carbon isotope ratios in clump stars of open clusters as a function of stellar TO mass. Red square indicates the value of IC\,4756. Red open squares indicate results of previously investigated open clusters by \cite{Tautvaisiene00, Tautvaisiene05, Tautvaisiene16, Mikolaitis10, Mikolaitis11a, Mikolaitis11b, Mikolaitis12, Drazdauskas16, Drazdauskas16b}. Other symbols include results from \cite{Gilroy89} -- pluses, \cite{Luck94} -- open circles, \cite{Smiljanic09} -- green crosses, \cite{Santrich13} -- open diamond. The solid lines (1DUP ST) represent the $^{12}{\rm C}/^{13}{\rm C}$ ratios predicted for stars at the first dredge-up with standard stellar evolutionary models of solar metallicity by \cite{Charbonnel10} (black solid line) and \cite{Lagarde12} (blue solid line). The short-dashed line (TH) shows the prediction when only thermohaline extra-mixing is introduced \citep{Charbonnel10}, and the long-dashed line (TH+V) is for the model that includes  both the thermohaline and rotation-induced mixing \citep{Lagarde12}. A typical error bar is indicated \citep{Charbonnel10, Smiljanic09, Gilroy89}.}
              \label{Fig. 12C13C}%
    \end{figure}
    
\begin{figure}
   \centering
        \includegraphics[width=\columnwidth]{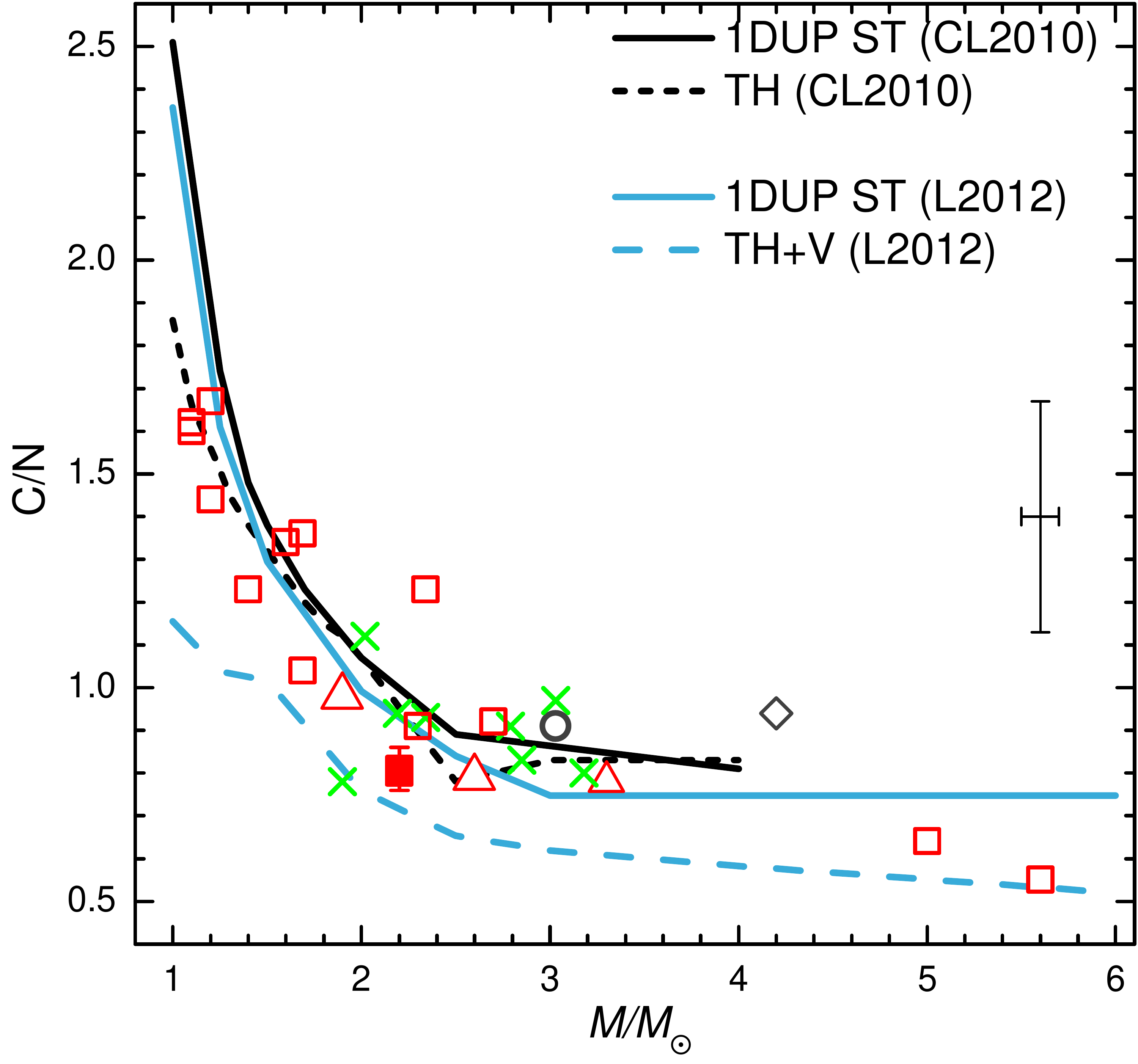}
   \caption{The average carbon-to-nitrogen ratios in clump stars of open clusters as a function of stellar TO mass. In addition to symbols in Fig. \ref{Fig. 12C13C}, here we include the results from \cite{Tautvaisiene15} as red open triangles.}
              \label{Fig. 12C14N}%
    \end{figure}

\begin{figure}
   \centering
        \includegraphics[width=\columnwidth]{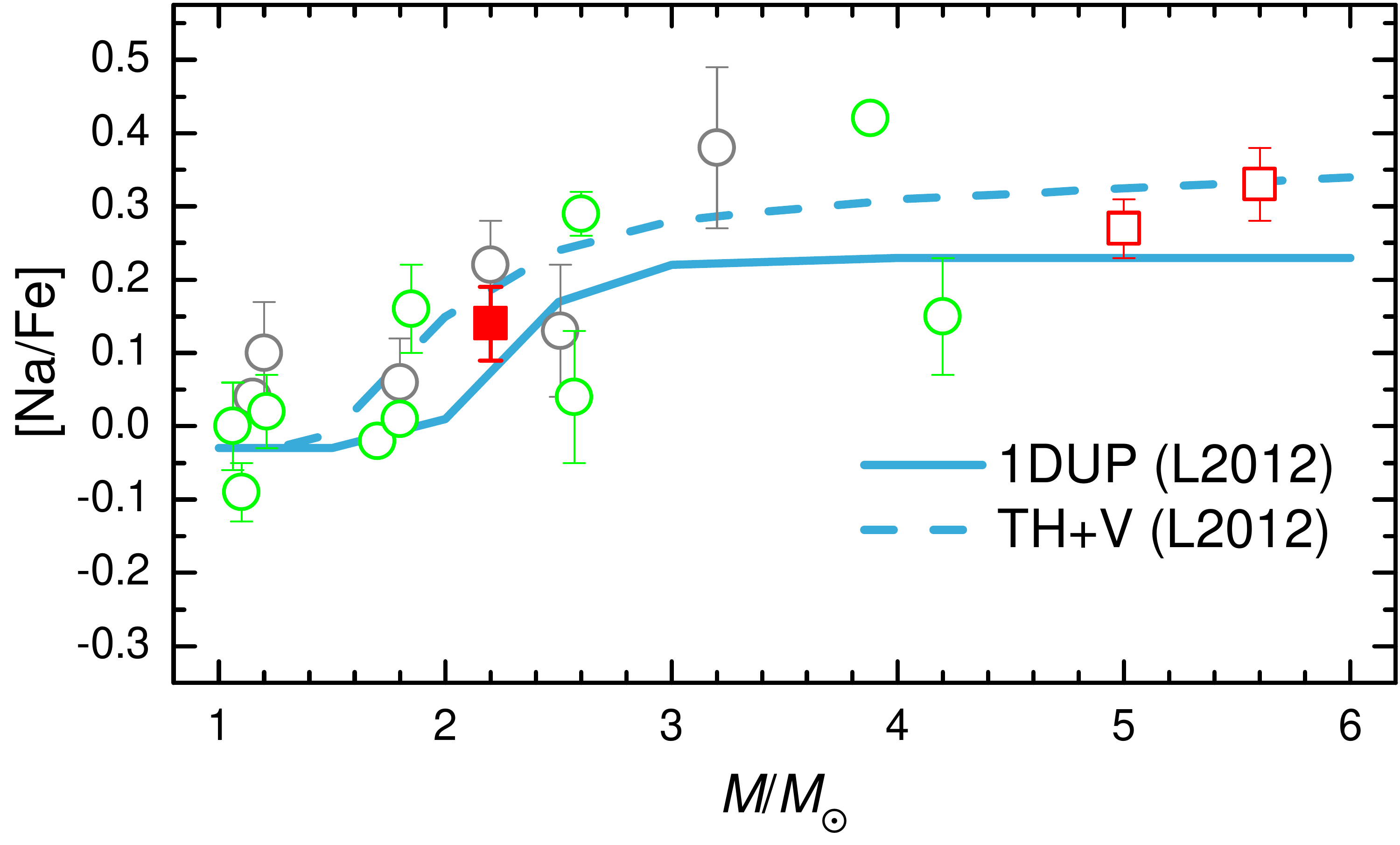}
   \caption{The mean [Na/Fe]$_{\rm NLTE}$ abundances in open clusters compared to theoretical models by \cite{Lagarde12}. The result obtained in this study 
   is marked with the red square. The red open squares indicate results from \cite{Drazdauskas16b}. The results from \cite{Maclean15} are indicated as the open grey circles and 
   from \cite{Smiljanic16} are shown as the green open circles.
   }
              \label{NaFe}
    \end{figure}

\begin{table*}
        \centering
    \caption{Comparison of mean IC\,4756 abundances determined in this work and previous studies.}
    \label{authorcomparison}
    \begin{tabular}{lrrrrrrr}
    \hline
    \hline
    \noalign{\smallskip}
        &       This study      &       Luck (1994)     & Jacobson et      & Smiljanic et              & Pace et al.     & Maiorca et       & Ting et al.  \\
Element /    &             &                   & al. (2007)  &   al. (2009)    & (2010)  &  al. (2011) & (2012)   \\
Resolution & 48\,000    &    18\,000        &   15\,000 &        48\,000        &     100\,000      &        100\,000   &    30\,000    \\
        \hline
        \hline
        \noalign{\smallskip}

{[C/Fe]}                        &       $\mathbf{-0.26\pm 0.03}$&       $-0.27\pm0.05$  &       ...     &       $\mathbf{-0.15\pm 0.02}$  & ... & ... &   ...     \\
{[N/Fe]}                        &       $\mathbf{0.45\pm 0.02}$ &       $0.54\pm0.16$   &       ...     &       $\mathbf{0.43\pm 0.07}$  &       ...     & ...&  ...     \\
{[O/Fe]}                        &       $\mathbf{0.01\pm 0.05}$ &       $-0.17\pm0.09$  &       ...     &       $\mathbf{-0.01\pm 0.01}$  &       $\mathbf{<0.07}$        & ...&  ...     \\
C/N                             &       $\mathbf{0.79\pm 0.05}$ &       $0.80\pm0.20$   &       ...     &       $\mathbf{1.05\pm 0.12}$  &       ...     & ...&  ...     \\
$^{12}{\rm C}/^{13}{\rm C}$     &       $\mathbf{19\pm 1.4}$    &       $\mathbf{15}$   &     ...     &       $\mathbf{13\pm 3.1}$    &       ...     & ...&  ...     \\
\noalign{\smallskip}
$\rm [Na/Fe]_{LTE}$             &       $0.22\pm0.05$           &       $0.14\pm0.01$   &       $0.57\pm0.06$   &       ...     &       $0.11\pm0.04$   & ...&    $0.21\pm0.15$   \\
$\rm [Na/Fe]_{NLTE}$        &   $0.14\pm0.05$           &       ...                 &    ...     &       $-0.01\pm0.04$  &       ...     & ...&  ...             \\
{[Mg/Fe]}                       &       $0.08\pm0.05$           &       $0.20\pm0.14$   &       ...     &       $-0.05\pm0.02$  &       ...     & ...&    $0.12\pm0.13$   \\
{[Al/Fe]}                       &       $0.04\pm0.03$           &       $0.07\pm0.21$   &       $0.29\pm0.08$   &       ...     &       $-0.11\pm0.05$  & ...&    $0.12\pm0.12$   \\
{[Si/Fe]}                       &       $0.04\pm0.03$           &       $0.23\pm0.04$   &       $0.34\pm0.06$   &       $0.06\pm0.04$   &       $0.02\pm0.01$   & ...&    $0.13\pm0.13$   \\
{[Ca/Fe]}                       &       $0.11\pm0.03$           &       $-0.07\pm0.13$  &       $0.07\pm0.08$   &       $0.02\pm0.04$   &       $-0.02\pm0.03$  & ...&    $0.05\pm0.16$   \\
{[Sc\,{\sc ii}/Fe]}             &       $-0.01\pm0.04$          &       $0.03\pm0.12$   &       ...     &       $0.06\pm0.07$   &       ...     & ...& ...        \\
{[Ti\,{\sc i}/Fe]}              &       $0.00\pm0.02$           &       $-0.09\pm0.13$  &       ...     &       $-0.04\pm0.01$  &       $0.03\pm0.03$   & ...& ... \\
{[Ti\,{\sc ii}/Fe]}             &       $-0.03\pm0.05$          &       ...                 &    ...     &       ...     &       ...     & ...&  $0.28\pm0.08$   \\
{[Fe\,{\sc i}/H]}               &       $-0.02\pm0.01$          &       $-0.05\pm0.05$  &      $-0.15\pm0.04$  &       $0.04\pm0.03$   &       $0.08\pm0.02$   & $0.01\pm0.03$ & $-0.01\pm0.10$  \\
{[Fe\,{\sc ii}/H]}              &       $-0.02\pm0.02$          &       $-0.04\pm0.04$  &      ...     &       $0.04\pm0.03$   &       ...     & ...&  $0.00\pm0.11$   \\
{[Cr/Fe]}                       &       $0.03\pm0.03$           &       $0.08\pm0.04$   &       ...     &       $0.04\pm0.05$   &       $0.00\pm0.03$   & ...&    $0.03\pm0.15$   \\
{[Mn/Fe]}                       &       $\mathbf{-0.07\pm0.05}$         &       $0.20\pm0.18$   &       ...     &       ...     &       ...     & ...&    ...     \\
{[Co/Fe]}                       &       $\mathbf{-0.04\pm0.02}$         &       $0.32\pm0.16$   &       ...     &       $0.06\pm0.04$   &       ...     & ...&    ...     \\
{[Ni/Fe]}                       &       $-0.06\pm0.01$          &       $0.08\pm0.07$   &       $0.08\pm0.05$   &       $-0.01\pm0.02$  &       $-0.04\pm0.01$  & ...&    $0.03\pm0.13$   \\
\noalign{\smallskip}
{[Y/Fe]}                        &       $\mathbf{0.03\pm0.04}$  &       $0.37\pm0.03$   &       ...     &       ...     &       ...     &  $0.11\pm0.01$  &       ...     \\
{[Zr/Fe]}                       &       $\mathbf{0.18\pm0.04}$  &       $0.25\pm0.13$   &       ...     &       ...     &       ...     & $0.09\pm0.02$ & ...     \\
{[Ba/Fe]}                       &       $\mathbf{0.23\pm0.04}$  &       $0.01\pm0.21$   &       ...     &       ...     &       ...     & ...&    $\mathbf{0.00\pm0.14}$  \\
{[La/Fe]}                       &       $\mathbf{0.22\pm0.04}$  &       ...                 &    ...     &       ...     &       ...     & $0.19\pm0.01$&        ...     \\
{[Ce/Fe]}                       &       $\mathbf{0.21\pm0.03}$  &       ...                 &    ...     &       ...     &       ...     & $0.16\pm0.02$&        ...     \\
{[Pr/Fe]}                       &       $\mathbf{0.18\pm0.04}$  &       ...                 &    ...     &       ...     &       ...     & ...&  ...     \\
{[Nd/Fe]}                       &       $\mathbf{0.12\pm0.04}$  &       $0.37\pm0.01$   &       ...     &       ...     &       ...     & ...&    ...     \\
{[Eu/Fe]}                       &       $\mathbf{0.08\pm0.04}$  &       $0.16$  &       ...     &       ...     &       ...     & ...&    ...     \\
    \noalign{\smallskip}
    \hline
    \end{tabular}
    \tablefoot{
    The abundances marked in bold were determined using spectral syntheses. The remaining abundances were derived from EWs. Along with the mean elemental abundances, values of root mean square between stars are presented.   
    }
\end{table*}

\subsection{$^{12}C/^{13}C$ and C/N ratios}

During a star's lifetime, complex nuclear reactions take place inside the core, producing different elements which, for the most part, stay deep inside stellar interiors. However, when the star leaves the main sequence and becomes a red giant, the convective envelope of the star deepens, and the processed material from the core is brought up to the surface. By looking at abundance alterations of certain elements in stellar photospheres, we can analyse the efficiency of the transport mechanisms. Among the most sensitive elements are carbon and nitrogen. 
$^{12}{\rm C}/^{13}{\rm C}$ and $^{12}{\rm C}/^{14}{\rm N}$ ratios are particularly good indicators of mixing in stars.  

Carbon and nitrogen abundances in evolved stars have been investigated for more than 45 years, since studies by \citet{Day73, Tomkin74, Tomkin75, Dearborn76} and others.
It is well known that stars undergo several mixing events during their evolution, however only one, the first dredge-up, was predicted by the classical theory of stellar evolution until a star leaves the red clump position on the HR diagram. As concerns studies of evolved stars in open clusters, inconsistencies between the $^{12}{\rm C}/^{13}{\rm C}$ ratios predicted by the classical stellar evolution model and the observational results were clearly demonstrated by \cite{Gilroy89}.  The observational results agreed relatively well with the classical model for stars in open clusters with turn-off (TO) masses larger than $\sim$~2.2~M$_{\odot}$. However, for stars in open clusters with lower turn-off masses the observed $^{12}{\rm C}/^{13}{\rm C}$ ratios were decreasing with decreasing TO mass.

In order to explain discrepancies between theoretical and observed abundances of mixing-sensitive chemical elements, some other transport mechanisms besides the classical convection had to be introduced. Nowadays, the most promising of these seems to be a thermohaline-induced extra-mixing. 
We compare our results with the first dredge-up model, a model with only the thermohaline mixing included \citep{Charbonnel10} and with a model where rotation and thermohaline-induced mixing act together 
\citep{Lagarde12}. Stellar rotation may modify an internal stellar structure even before the RGB phase, however results become visible only in later stages. 
The initial rotation velocity of the models on the zero age main sequence (ZAMS) was chosen at 30\% of the critical velocity at that point
and leads to the mean velocity of about 120~km\,s$^{-1}$ \citep{Lagarde14}. We present the comparison of our results and the models in Figs.~\ref{Fig. 12C13C} and \ref{Fig. 12C14N}.
 The TO mass of IC\,4756 was determined using PARSEC isochrones \citep{Bressan12}. The input metallicity of $-0.02$ dex was the one we determined in this work, and we took the cluster age of 0.89~Gyr as determined by \cite{Strassmeier15}. We consider all the stars in our analysis as being in a He-core burning stage. Our determined TO mass of IC\,4756 is around 2.2~$M_{\odot}$. 
 
 The average $^{12}{\rm C}/^{13}{\rm C}$ value as obtained from three evolved IC\,4756 giants is $19 \pm 1.4$ and the mean C/N ratio is equal to $0.79\pm 0.05$. The previous analysis of C/N ratios were made by \cite{Luck94} and \cite{Smiljanic09}. Our result agrees well with the one by \cite{Luck94} however the mean C/N value reported by \cite{Smiljanic09} is slightly larger and has a larger scatter (see Table~\ref{authorcomparison}). From the same stars, \cite{Smiljanic09} obtained a somewhat lower result for $^{12}{\rm C}/^{13}{\rm C}$ and a larger scatter. The \cite{Luck94} result lies in the middle with a result from only one star with the value of 15. The mean $^{12}{\rm C}/^{13}{\rm C}$ ratio determined for this cluster by \cite{Gilroy89} agrees well with our determination.  
 
 From the comparison of our result and the theoretical models we can see that the mean values of $^{12}{\rm C}/^{13}{\rm C}$ and C/N ratios lie between the model 
 with only the thermohaline extra-mixing included and the model which also includes the rotation-induced mixing. The rotation was most probably smaller 
 in the investigated IC\,4756 stars than 120~km\,s$^{-1}$ when they were on the ZAMS. 
When looking at the C/N values from ours and previous studies, it is obvious that the pure thermohaline extra-mixing model is preferred for the open clusters with the TO masses below 2.2~$M_{\odot}$. 
There were some debates (see \citealt{Wachlin14} and references therein) that thermohaline convection models greatly overestimate the effects of such mixing, and some other explanation might be needed. 

\subsection{Sodium}

\begin{figure}
   \centering
        \includegraphics[width=\columnwidth]{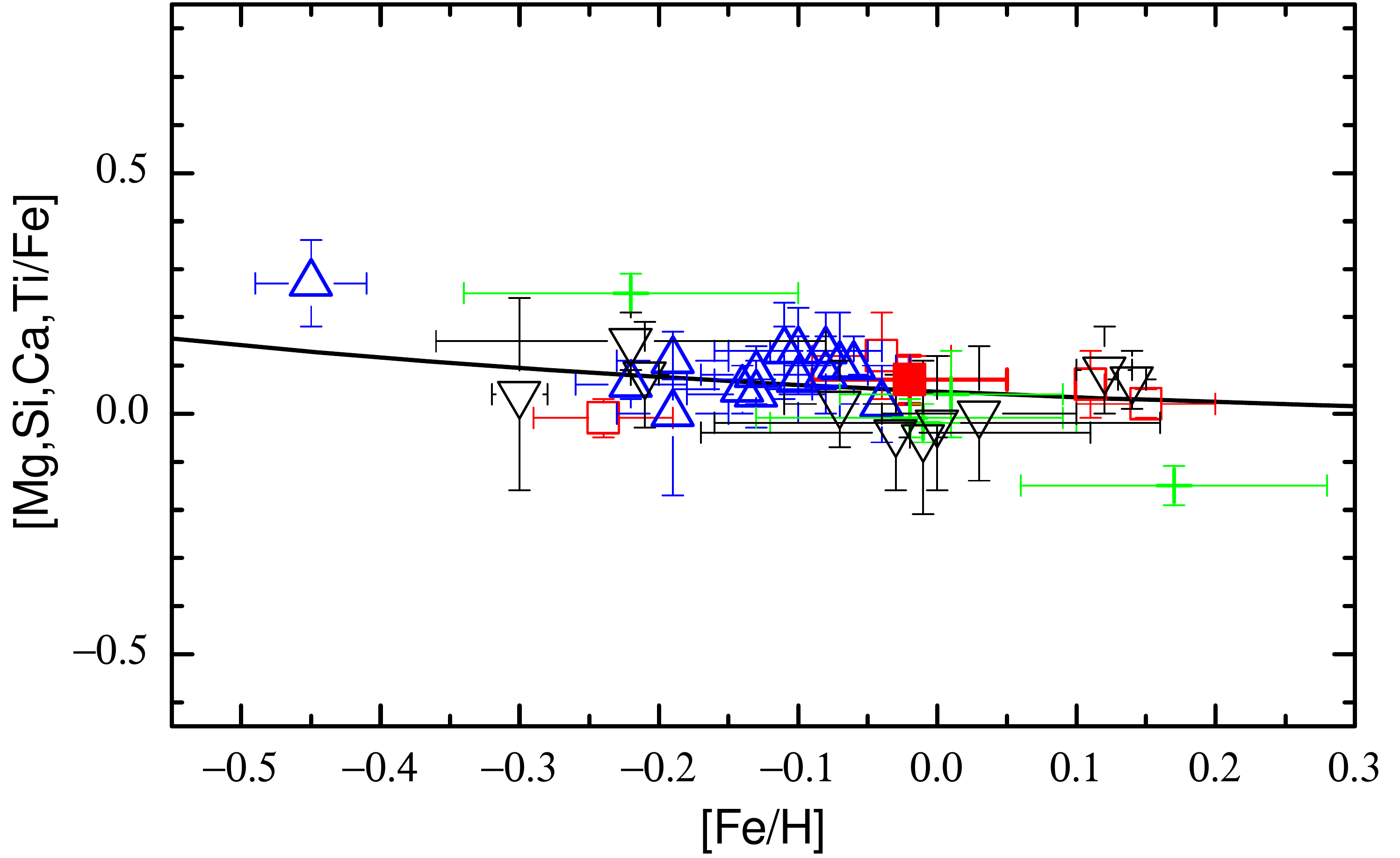}
   \caption{Mean $\alpha$-element abundances in open cluster RGB stars. The result for this study is indicated by the red circle. Results from \cite{Tautvaisiene05, Mikolaitis10, Mikolaitis11a, Mikolaitis11b, Drazdauskas16b} are marked as the red open squares. The blue triangles indicate results from \cite{Reddy12, Reddy13, Reddy15}; the green plus signs from \cite{Mishenina15}; and the black reverse triangles from \cite{Friel10, Jacobson08, Jacobson09}. The black line represents the Galactic disc evolution model by \cite{Pagel95}.}
              \label{alpha_average}
\end{figure}

\begin{figure}
   \centering
        \includegraphics[width=\columnwidth]{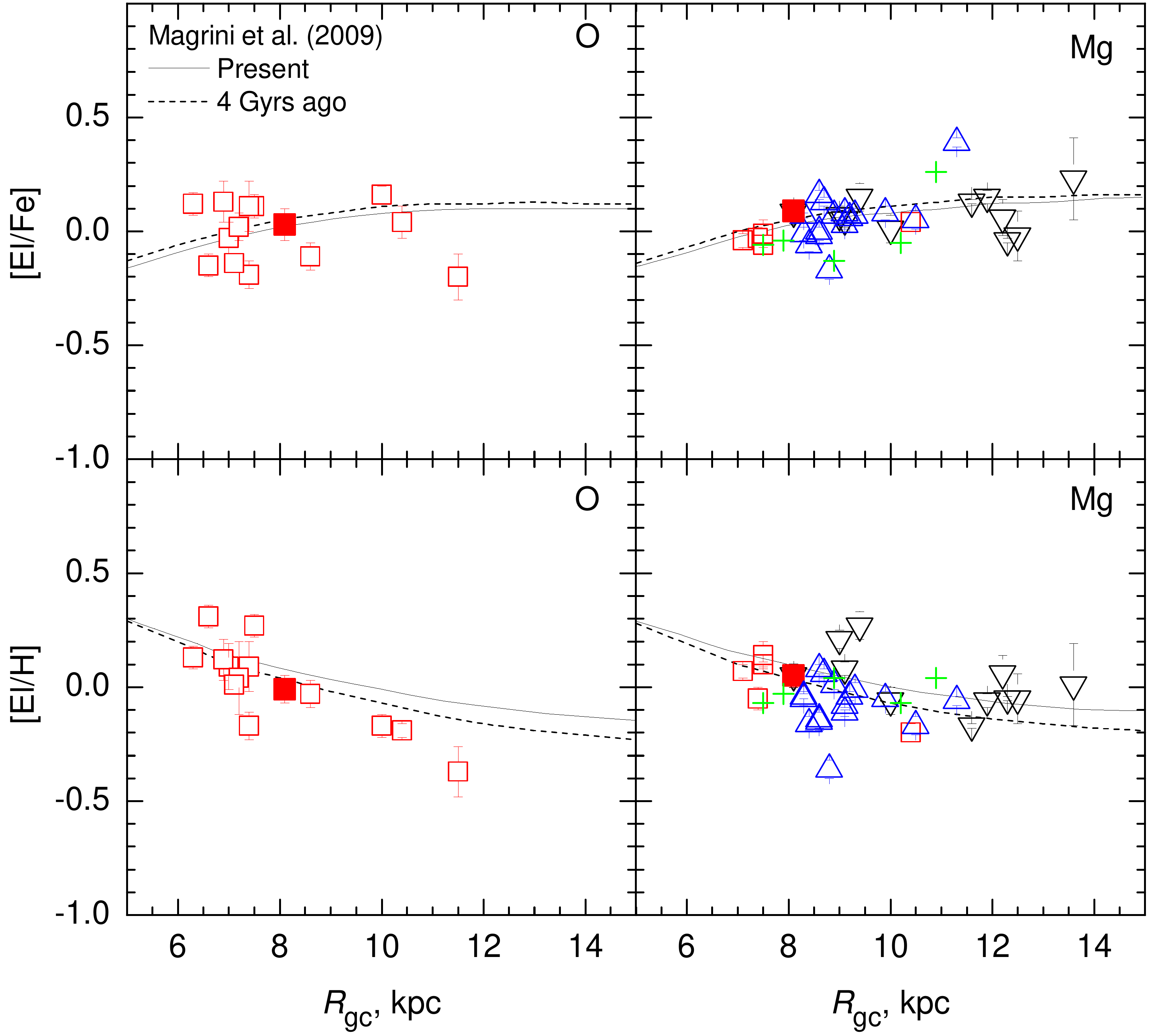}
   \caption{Mean oxygen and magnesium abundances compared to the theoretical models by \cite{Magrini09}. The oxygen results are taken from \cite{Tautvaisiene05, Mikolaitis10, Mikolaitis11a, Mikolaitis11b, Tautvaisiene15, Drazdauskas16, Drazdauskas16b}. Symbols are the same as in Fig.~\ref{alpha_average}.}
              \label{oxygen_magnesium}
\end{figure}

Sodium, besides carbon and nitrogen, is one of the other mixing-sensitive chemical elements. Theoretical models by \cite{Lagarde12} predict a significant increase of sodium abundance after the first dredge-up for stars with turn-off masses of around 2~M$_{\odot}$, and an even larger overabundance if the extra-mixing is taken into account. 
Our results and several other recent NLTE determinations of sodium abundances in other studies (\citealt{Drazdauskas16b, Maclean15}, and \citealt{Smiljanic16}) are displayed in Fig.~\ref{NaFe} together with the theoretical models by \citep{Lagarde12}. Following \citet{Smiljanic16}, only high-probability giant stars were extracted from the paper by \cite{Maclean15}. The theoretical models show how sodium abundance increases in relation to the TO mass. 
At the TO mass of IC\,4756, which is around 2.2~$M_{\odot}$, the average NLTE sodium abundance, as in case of $^{12}{\rm C}/^{13}{\rm C}$ and C/N, lies between the theoretical model which includes the thermohaline- and rotation-induced extra-mixing and the model of the first dredge-up (or the model of pure thermohaline mixing since they both give very close [Na/Fe] values). These results confirm a conclusion by other investigators who suggest that the trend of sodium abundance increases in relation to a TO mass and is most probably caused by internal stellar evolutionary processes (\citealt{Smiljanic16} and references therein).

\subsection{$\alpha$-elements}

$\alpha$-elements are of particular interest when studying Galactic archaeology due to a different timescale of their production compared to iron-peak elements. These elements are mainly produced in massive stars, thus any visible enhancement in their abundances can reveal differences of star formation histories in different parts of the Galaxy.

Figure~\ref{alpha_average} displays the determined average $\alpha$-element abundance in IC~4756 together with results from other studies. The average abundances were calculated simply by taking the average of [Mg/Fe], [Ca/Fe], [Si/Fe], and [Ti/Fe] values. We find no $\alpha$-element enhancement in our cluster, and our result shows an almost perfect match with the theoretical model by \cite{Pagel97}.

We take a look at oxygen and magnesium abundances separately in Fig.~\ref{oxygen_magnesium} and compare them with models by \cite{Magrini09} and abundance results from other studies. Oxygen is one of the more difficult elements to precisely analyse due to the relations with carbon and nitrogen. Therefore, for the comparison we take the results from studies where the oxygen abundance was determined in the same way as in this study \citep{Mikolaitis10, Mikolaitis11a, Mikolaitis11b, Tautvaisiene15, Drazdauskas16, Drazdauskas16b}. For the magnesium, we have taken results from the same authors as other $\alpha$-elements \citep{ Mikolaitis10, Mikolaitis11a, Mikolaitis11b, Reddy12, Reddy13, Reddy15, Mishenina15, Friel10, Jacobson08, Jacobson09}. The models depict the oxygen and magnesium abundance trends in relation to the Galactocentric distance. Our results for IC\,4756 follow the models, and at $R_{\rm gc}$ of 8.1~kpc show no visible over- or under-abundances for oxygen or magnesium.

\subsection{Iron-peak elements}

The mean abundances of five iron-peak elements investigated in our sample of IC\,4756 RGB stars display no significant deviations from the mean iron abundance.  These results are consistent with previous abundance determinations for this cluster (see Table~\ref{authorcomparison}). 
 Larger discrepancies are coming from the study by \cite{Luck94} who reported relatively large values and scatters for manganese and cobalt, which might be caused by uncertainties in accounting for a hyperfine structure in these element lines.

\subsection{Neutron-capture elements}

Along with our results, in Table~\ref{authorcomparison} we present abundances of neutron-capture elements determined in IC\,4756 by several other studies. The results by \citet{Luck94} were averaged from one supergiant and two giants.  \citet{Maiorca11} investigated three dwarfs using an EW method, while \citet{Ting12} from 12 giant stars (10 of which are common with our work) determined solely the barium abundance.

\begin{figure}
    \centering
        \includegraphics[width=\columnwidth]{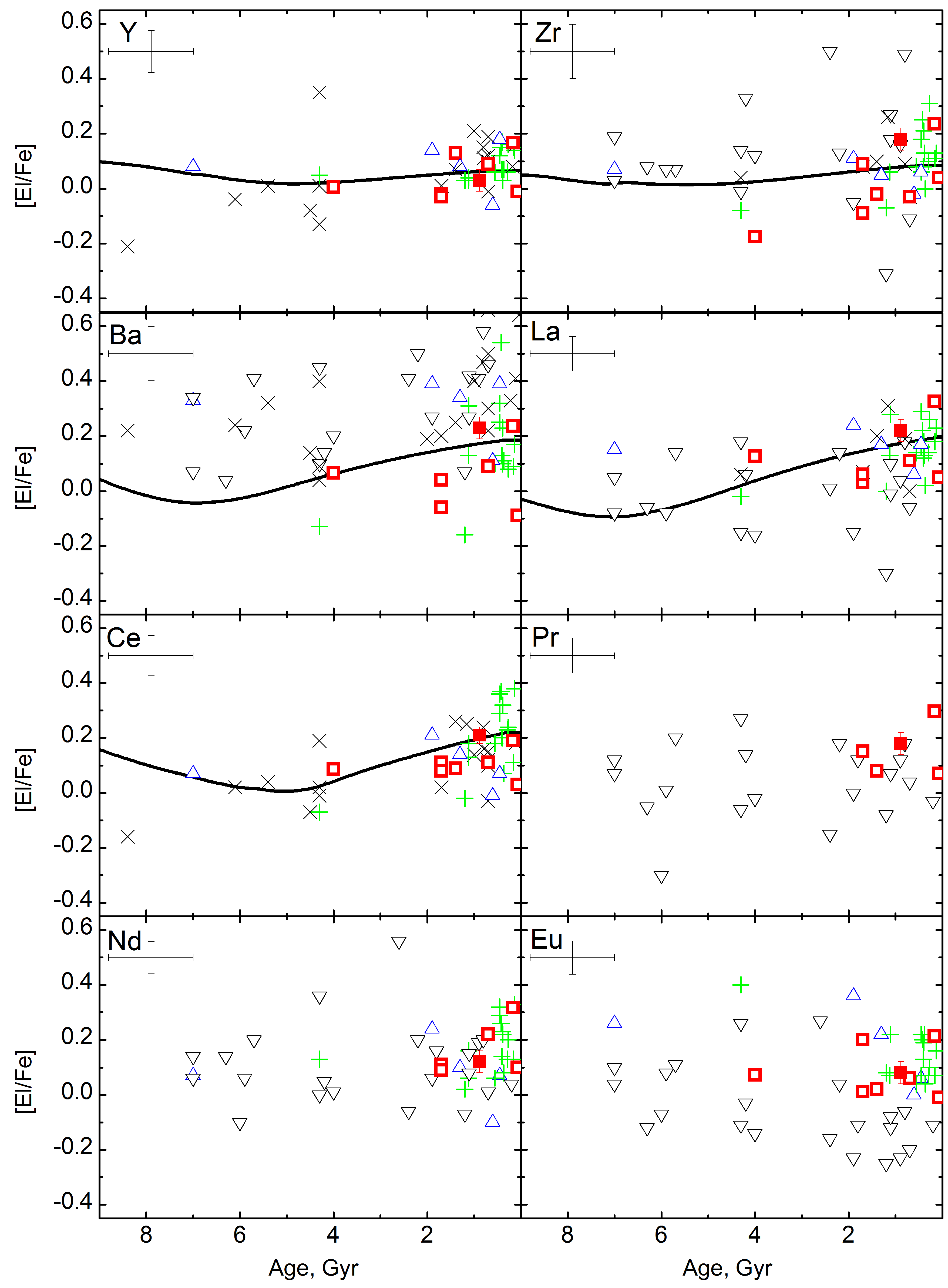}
        \caption{Averaged values of neutron-capture element abundances vs. age in IC\,4756 (red filled squares) and other previously investigated clusters. The green plus signs indicate results by \citet{Reddy12, Reddy13, Reddy15}; the blue triangles -- results by \citet{Mishenina15}; the black crosses -- results by \citet{D'Orazi09} and \citet{Maiorca11}; the black reverse triangles -- results by \citet{Jacobson13} and \citet{Overbeek16}; the empty red squares -- results from our previous studies by \citet{Tautvaisiene00, Tautvaisiene05}, \citet{Mikolaitis10, Mikolaitis11a, Mikolaitis11b} and \citet{Drazdauskas16}. The adopted approximate age errors of 0.9~Gyr are based on determinations by \citep{Salaris04}. The continuous lines indicate a chemical evolution model by \citet{Maiorca12} at the Solar radius.
    }
    \label{neutron_capture_age}
\end{figure}

Y and Zr could be attributed to the elements that lie in the first $s$-process peak and are referred to as the light $s$-process elements. Their production scenarios are similar. In the Sun, 74\% of yttrium and 67\% of zirconium are produced via the $s$-process \citep{Travaglio04}. The main $s$-process component, which gathers free neutrons from low-mass AGB stars, contributes to 69\% of Y and to 65\% of Zr, while the weak $s$-process component contributes only 5\% of Y and 2\% of Zr. The rapid neutron capture process ($r$-process) is responsible for 8\% of Y and 15\% of Zr. A responsible process for production of the remaining 18\% of Y and Zr is debatable. \citet{Travaglio04} called this process the lighter element primary process (LEPP) which possibly acts in low-metallicity massive stars. Therefore, due to the similarities of production of Y and Zr, abundances of these elements should be similar. As displayed in  Fig.~\ref{neutron_capture_age}, the clusters' results from various studies of
 Y and Zr exhibit abundances roughly confined between 0 and 0.2~dex, with several values scattered around this range. Our programme cluster IC\,4756 displays average abundance ratios of 0.03~dex for [Y/Fe], and 0.18~dex for [Zr/Fe]. Abundances of those elements are not similar, however, they follow the modeled trends by \citet{Maiorca12} quite well. 

Ba, La, and Ce belong to the second s-process peak. According to \citet{Arlandini99}, 
$s$-process contributes to Ba, La, and Ce production by 81\%, 62\%, and 77\%, respectively. The newer study by \citet{Bisterzo16} provided slightly different results, indicating the $s$-process enrichment of Ba, La, and Ce by 83\%, 73\%, and 81\%, respectively. This strengthens a key feature of the elements in the second peak that all elements beyond $A=90$, except for Pb, are mainly produced via the main $s$-process component \citep{Kappeler11}.  The remaining abundances of these elements are provided by the $r$-process. Thus, the similarity of the Ba, La and Ce production should be reflected in their abundance similarities. By looking at Table~\ref{authorcomparison}, we see that the average abundances of these elements are identical in our work.

Nd and Pr are elements that have a mixed origin, as they are produced via s- and r-processes in roughly equal fractions. As reported by \citet{Arlandini99} and \citet{Bisterzo16}, $s$-process produces around 49\% of Pr and 56\% of Nd, while the remaining fractions of the elements are created via the $r$-process – 51\% and 44\%, respectively. Rather similar abundances of Nd and Pr in IC\,4756 confirm their formation scenario.

Finally, europium is an almost purely $r$-process-dominated element, as only 6\% of its production takes place via the $s$-process and the remaining 94\% via the $r$-process \citep{Bisterzo16}. The europium abundance in IC\,4756 is similar to the thin disc chemical content (e.g. \citealt{Pagel97}).

From the comparison of n-capture chemical element abundances in open clusters and a semi-empirical models of the Galactic thin disc chemical evolution by \citet{Pagel97}, we see that s-process-dominated elements, and especially barium, have higher abundances for many open clusters. This phenomenon can be explained by the young age of these clusters and a larger contribution of low-mass asymptotic giant branch stars in producing s-process elements at the time these clusters were formed (c.f. \citealt{D'Orazi09}, \citet{Maiorca11}). At the same time, the r-process-dominated element europium results agree well with the thin disc model.

In Fig.~\ref{neutron_capture_age} we can see how abundances of n-capture element abundances in open clusters depend on cluster ages. IC\,4756 is a relatively young open cluster with an age of around 0.8~Gyr. Abundances of all investigated elements in IC\,4756 agree well with the model by \citet{Maiorca12} for the Solar radius. This is applicable also to barium (nine IC\,4756 giants provide the average [Ba/Fe]$= 0.22\pm0.03$), while [Ba/Fe] in OC investigated in several other studies at similar ages have values from 0.3~dex to 0.6~dex exceeding predictions of the model (see Fig.~\ref{neutron_capture_age}).

\section{Conclusions}
Recognising the role that open clusters play in the establishment of Galactic chemical evolution models, we performed a detailed high-resolution spectroscopic analysis of the open cluster IC\,4756. We determined the main atmospheric parameters of 13 giant stars, with 9 of them being the high-probability members of the cluster. We also carried out a comprehensive analysis of 23 chemical elements made in various stellar phases and processes.
The key results of our analysis are as follows:

\begin{itemize}
    \item IC\,4756 has a metallicity close to Solar. ${\rm [Fe/H]}= -0.02\pm 0.01$, whereas other iron-peak 
    elements do not differ by more than 0.1~dex from Solar values.
    \item Our determined average $\alpha$-element abundances show a slight enrichment of 0.07~dex compared to iron, 
    however that is expected from the thin disc chemical evolution model by \citet{Pagel95}. 
    The results for oxygen and magnesium show no deviation from the models by \citet{Magrini09}, also at the given galactocentric distance of 8.1~kpc.
    \item  The mean ratio of carbon and nitrogen,  C/N = $0.79\pm0.05$, and the carbon isotope ratio, 
    $^{12}{\rm C}/^{13}{\rm C}$ = $19\pm1.41$, are altered more than predicted in the first dredge-up model, 
    and lie between the model where only the thermohaline extra-mixing is included and the model which also 
    includes the rotation-induced mixing (\citealt{Lagarde12}). The value 
    we obtained for sodium with NLTE corrections, ${\rm [Na/Fe]}=0.14\pm0.05$, is larger than the first dredge-up prediction, however this is  
    lower than the value predicted by the model of thermohaline- and rotation-induced extra mixing. The rotation was most probably 
    smaller in IC\,4756 stars than the 30\% of the critical rotation velocity at the ZAMS.
    \item Being relatively young, the open cluster IC\,4756 displays a moderate enrichment of $s$-process-dominated chemical elements compared to 
    the Galactic thin disc model (\citealt{Pagel97}) and confirms the enrichment of $s$-process-dominated elements in young open clusters compared 
    to the older ones. Abundances of all investigated $s$-process-dominated elements in IC\,4756 agree well with the model for the Solar radius by \citet{Maiorca12}.
    The $r$-process-dominated element europium abundance agrees with the thin disc abundance.
\end{itemize}

\begin{acknowledgements}
This research has made use of the WEBDA database (operated at the Department of Theoretical Physics and Astrophysics of the Masaryk University, Brno), of SIMBAD (operated at CDS, Strasbourg), of VALD (\citet{Kupka00}), and of NASA's Astrophysics Data System. Bertrand Plez (University of Montpellier II) and Guillermo Gonzalez (Washington State University) were particularly generous in providing us with atomic data for CN and C$_2$, respectively. We thank Laura Magrini for sharing with us the Galactic chemical evolution models. We also thank the referee for helpful suggestions
and comments that improved the quality of this paper. VB, AD, GT, YC were partially supported by the grant from the Research Council of Lithuania (MIP-082/2015). RS acknowledges support by the National Science Center of Poland through the grant 2012/07/B/ST9/04428.

\end{acknowledgements}

%
   \bibliographystyle{aa} 
   \bibliography{References3} 
%

\end{document}